\documentclass[aps,prl,groupedaddress,twocolumn,superscriptaddress,amsfonts,amssymb,amsmath,showpacs,floatfix,preprintnumbers]{revtex4-2}

\usepackage{dcolumn}
\usepackage{bm}
\usepackage{multirow}
\usepackage{verbatim}
\usepackage{graphicx}
\usepackage{ulem}
\usepackage{color}
\usepackage{placeins}
\usepackage{xcolor}
\usepackage{hyperref}
\usepackage{slashed}

\begin{document}

\preprint{JLAB-THY-25-4347}

\title{Window observables for benchmarking parton distribution functions}


 
\author{Joe Karpie}
\email[e-mail: ]{jkarpie@jlab.org}
\affiliation{Thomas Jefferson National Accelerator Facility, Newport News, VA 23606, U.S.A.}
\author{Christopher J.~Monahan}
\email[e-mail: ]{cjmonahan@coloradocollege.edu}
\affiliation{Department of Physics, Colorado College, Colorado Springs, CO 80903, U.S.A.}
\author{Kostas Orginos}
\email[e-mail: ]{kostas@wm.edu}
\affiliation{Physics Department, William \& Mary, Williamsburg, VA 23187, U.S.A.}\affiliation{Thomas Jefferson National Accelerator Facility, Newport News, VA 23606, U.S.A.}
\author{Savvas Zafeiropoulos}
\email[e-mail: ]{savvas.zafeiropoulos@cpt.univ-mrs.fr }
\affiliation{Aix Marseille Univ, Universit\'e de Toulon, CNRS, CPT, Marseille, France}

\begin{abstract}
Global analysis of collider and fixed-target experimental data and calculations from lattice quantum chromodynamics (QCD) are used to gain complementary information on the structure of hadrons. We propose novel ``window observables'' that allow for higher precision cross-validation between the different approaches, a critical step for studies that wish to combine the datasets. Global analyses are limited by the kinematic regions accessible to experiment, particularly in a range of Bjorken-$x$, and lattice QCD calculations also have limitations requiring extrapolations to obtain the parton distributions. We provide two different ``window observables'' that can be defined within a region of $x$ where extrapolations and interpolations in global analyses remain reliable and where lattice QCD results retain sensitivity and precision.
\end{abstract}

\maketitle

\section{Introduction}
Lattice quantum chromodynamics (QCD) is a vital tool for the precision study of particle physics, from quark flavor physics~\cite{FlavourLatticeAveragingGroupFLAG:2024oxs} to the muon's anomalous magnetic moment~\cite{Aoyama:2020ynm}. Recent theoretical and computational developments have also enabled lattice QCD calculations to become increasingly important in the quest to understand the internal structure of hadrons.

Parton distribution functions (PDFs) encode the longitudinal momentum structure of hadrons. The field-theoretic definition of PDFs relies on the hadronic matrix elements of fields at light-like separations, which cannot be determined directly in Euclidean lattice QCD and must be extracted using indirect approaches. The primary methods for determining PDFs from lattice QCD are based on Mellin moment calculations of matrix elements of local operators or inference of PDFs from matrix elements of non-local operators, through short-distance factorization (SDF)~\cite{Braun:2007wv,Radyushkin:2017cyf,Ma:2017pxb} or large momentum effective theory (LaMET)~\cite{Ji:2013dva}. 

Mellin moments correspond to weighted integrals over PDFs, so a potential challenge for robust comparison between phenomenological determinations of PDFs and the low Mellin moments determined from lattice calculations is that experimental data do not cover the entire range of Bjorken-$x$. 
A major goal of both the future Electron-Ion Collider~\cite{Accardi:2012qut} and the JLab 22~GeV upgrade~\cite{Accardi:2023chb} is to extend the range of the available data to reduce the extrapolation errors in PDFs, including the challenging case of nucleon quark transversity.

To determine PDFs via SDF or LaMET, non-local matrix elements calculated on the lattice must be Fourier transformed, which leads to an inverse problem associated with the incomplete Fourier transform of discrete, noisy data. Various approaches have been proposed to tackle this inverse problem
~\cite{Karpie:2019eiq,Liang:2019frk,DelDebbio:2020rgv,Cichy:2019ebf,Khan:2022vot,Chowdhury:2024ymm,Alexandrou:2020tqq,Candido:2024hjt,Dutrieux:2024rem,Dutrieux:2025jed,Karpie:2021pap}. Quantification of the systematic uncertainties associated with each approach is a significant difficulty and requires validation through multiple lattice calculations using various methods.

As a result of this inverse problem, lattice calculations that apply the SDF and LaMET approaches are most sensitive to the larger $x$ region where experimental data may be lacking. The PDF reconstructed from lattice results is highly correlated point-by-point in $x$ and therefore integrals over $x$ may have significantly higher precision than anticipated from the point-by-point variance. Inspired by the successful use of windows for the hadronic vacuum polarization contribution to the muon's anomalous magnetic moment~\cite{RBC:2018dos}, these difficulties motivate the use of ``window observables'' defined across a limited region of $x$ from $x_-$ to $x_+$. 

Benchmarking and validation of lattice calculations have been an important feature in flavor physics for two decades~\cite{FlavourLatticeAveragingGroupFLAG:2024oxs} and is a necessary step for meaningful comparison between lattice results and experimental determinations of hadron structure~\cite{Lin:2017snn,Constantinou:2020hdm}. Window observables provide benchmark quantities that enable quantitative comparison between, and validation of, lattice calculations of $x$-dependent hadron structure for three reasons.  First, window observables enable direct comparison between PDFs extracted from SDF and LaMET, and can be calculated across regions of momentum fraction or Ioffe time in which SDF and LaMET are expected to be reliable. Second, window observables provide quantities that can be extracted directly from phenomenological determinations of PDFs over regions of momentum fraction for which the PDFs are well constrained. These quantities can then be compared to window moments determined from SDF and LaMET. Third, window observables account for the large point-by-point correlations in PDFs and provide more precise benchmark quantities than a pointwise comparison between PDFs extracted from different approaches. Even in low $x$ windows, where experimental data exists and lattice results are noisy, the window observables provide meaningful benchmarks, as we demonstrate through the example of the isovector $u-d$ quark transversity.

This work is organized as follows. We briefly review parton structure and then propose two benchmark window observables for meaningful quantitative comparison between SDF, LaMET, and global analyses: window moments and Gaussian windows. To demonstrate a practical application of these window observables, we study synthetic data for the isovector $u-d$ quark transversity of the nucleon determined from JAM3D$\ast$~\cite{Gamberg:2022kdb,Cocuzza:2023oam} and JAMDiFF~\cite{Cocuzza:2023vqs,Cocuzza:2023oam}. These data enable closure tests for the window observables in the SDF framework.
We then apply the same procedure to lattice QCD results from~\cite{HadStruc:2021qdf}. Finally, we summarize and give future prospects. 

\section{Parton Structure\label{sec:definitions}}

Parton structure can be extracted from quantities defined in three interrelated spaces, Ioffe time, momentum fraction, and Mellin, each with advantages and disadvantages. The three representations provide complementary approaches to analyzing hadron structure. 
The LaMET~\cite{Ji:2013dva} and SDF~\cite{Radyushkin:2017cyf} approaches start from space-like Ioffe Time Distributions (ITDs) $\mathcal{M}(\nu,z^2)$, which were studied earlier in the context of ``primordial TMDs''~\cite{Musch:2010ka}. Experimental processes, described with Bjorken-$x$ and other momentum fractions, are most naturally analyzed with the PDF. Finally, direct lattice QCD calculations of matrix elements of local operators correspond to integer Mellin moments.

\paragraph{Ioffe-time representation}
The natural field-theoretic starting point is the renormalized matrix element, dubbed the ITD~\cite{Braun:1994jq},
\begin{equation}
    M(\nu,\mu^2) = \frac{1}{n\cdot p}\langle p | \bar{\psi}(z) \slashed{n} W(z;0) \psi(0)|p\rangle_{\mu^2} \,,
\end{equation}
with the quark fields connected by a straight Wilson line $W$ along the light-like separation $z_\mu=(0_+, z_-, 0_T)=z_- n$, with $n^2=0$. The ITD is a function of the Ioffe time $\nu=-p\cdot z$. In the rest frame of the hadron, such as in a fixed-target experiment like deep inelastic scattering, the Ioffe time $\nu=-mt$ is proportional to the probe interaction time, in units of the hadron mass~\cite{Ioffe:1969kf,IOFFE1969123}.

Within the SDF approach, the space-like ITD $\mathcal{M}(\nu,z^2)$, with $z_\mu=(0,0,0,z_3) = z_3 n$ and $n^2=-1$, is referred to as the pseudo-ITD. The pseudo-ITD is related to the ITD through~\cite{Radyushkin:2017cyf}
\begin{equation}
\mathcal{M}(\nu, z^2) =  \int_0^1 du\, C(u,\mu^2, z^2) M(u \nu, \mu^2) +O(z^2),\label{eq:pseudo}
\end{equation}
where in the parton limit $C(u)=\delta(1-u)$. The matching kernels have been calculated to $O(\alpha_S)$~\cite{Radyushkin:2018cvn,Zhang:2018ggy,Izubuchi:2018srq} and $O(\alpha_S^2)$~\cite{Li:2020xml}.

\paragraph{Momentum fraction representation}

The PDF, which arises naturally through QCD factorization of select scattering processes~\cite{Collins_book}, is defined as the inverse Fourier transform of the ITD
\begin{equation}
    f(x,\mu^2) = \int_{-\infty}^\infty \frac{d\nu}{2\pi} e^{- i\nu x} M(\nu,\mu^2)\,,
\end{equation}
and can be obtained from the pseudo-ITD through~\cite{Radyushkin:2017cyf,Ma:2017pxb}
\begin{equation}
    \mathcal{M}(\nu, z^2) = \int_{-1}^1 dx \,K(x\nu, \mu^2, z^2) f(x,\mu^2) +O(z^2)\,.
\end{equation}
The PDF can also be extracted in the LaMET approach through a related matching relation~\cite{Ji:2013dva,Ji:2020ect}.

\paragraph{Mellin representation}
The Operator Product Expansion (OPE) of the operator defining the ITD~\cite{Collins_book} is
\begin{equation}
    \bar{\psi}(z) \slashed{n} W(z;0) \psi(0) =\sum_n z_{\mu_1}\dots z_{\mu_n} \bar{\psi}\gamma^{\{\mu_1}D^{\mu_2}\dots D^{\mu_n\}}\psi
\end{equation}
where $\{\dots\}$ represents a traceless and symmetric combination of indices. The ITD itself becomes a series in Ioffe time
\begin{equation}
    M(\nu,\mu^2) = \sum_{n=0}^\infty \frac{(i\nu)^n}{n!} a_{n+1}(\mu^2)\,,
\end{equation}
where $a_n(\mu^2)$ are the reduced matrix elements of the local operators in the OPE, which are directly related to the Mellin moments of the PDF through 
\begin{equation}
a_n(\mu^2) = \int_{-1}^1 dx\, x^{n-1} f(x,\mu^2)\,.
\end{equation}
Alternatively, one can calculate the Mellin moments of the pseudo-PDF from polynomials of $\nu$~\cite{Karpie:2018zaz,Pang:2024kza} from a similar OPE for the pseudo-ITD~\cite{Ma:2017pxb}. 

In the rest of this paper, we suppress the renormalization scale dependence in all quantities.

\paragraph{Window observables}
In this work, we study convolutions of the PDF that constrain the information to a limited region of $x$. Generically, these are given by
\begin{equation}
W_n(x_\pm) = \int_{-1}^1 dx\,w_n(x;x_\pm) f(x)\,.
\end{equation}
In both examples below, as the magnitude of $|n|$ increases, up to some normalization, the window observables are dominated by the PDFs at a single point $x=x_+$, $x_-$, or their average. At smaller $|n|$ they represent relatively wider features within the prescribed window $(x_-, x_+)$, and lattice QCD reconstructions will be more precise than the large $|n|$ approximate point-by-point reconstruction.

One example is ``window moments''
\begin{equation}
    a_n(x_-,x_+) = \int_{x_-}^{x_+} dx\, x^n f(x) \,.
\end{equation}
The related ``truncated moments'', defined with $x_+=1$, were first introduced in~\cite{Forte:1998nw,Forte:2000wh} and further studied in~\cite{Piccione:2001vf,Kotlorz:2006dj,Psaker:2008ju,Aschenauer:2012ve,Kotlorz:2014kfa,Kotlorz:2014fia,Kotlorz:2016icu}. We focus on moments with positive $n$, which lose sensitivity to the lower region as $n$ grows. Moments with $n$ negative have decreasing sensitivity to the upper region and give less reliable lattice QCD benchmarks.

We also propose an alternative window observable, which we call ``Gaussian windows'', defined by the PDF convoluted with a Gaussian localized in $x$,
\begin{equation}
    g_n(x_-,x_+) = \int_{-1}^{1} dx \sqrt{\frac{n^2}{2\pi x_d^2}} \exp\Big[ -n^2 \frac{(x-x_0)^2}{2 x_d^2}\Big] f(x) \,,\label{eq:gauss_window_def}
\end{equation}
where $x_0=(x_++x_-)/2$ and $x_d=x_+-x_-$. Increasing $|n|$ tightens the window about the midpoint $x_0$. As $|n|\to\infty$, the Gaussian window approaches $f(x_0)$. In some cases, such as the analysis of generalized parton distributions (GPDs), which exhibit a complicated analytic structure at $x=\xi$, it may be useful to work with locally smeared observables like a Gaussian window with $x_0\approx\xi$ rather than exact point-by-point values.

Both window moments and Gaussian windows can be obtained from direct integration over $x$ or, as described in the Supplementary Materials, the convolution theorem can be used to recast window observables as integrations in $\nu$. The window observables therefore provide an environment in which direct quantitative comparisons between SDF, LaMET, and global phenomenological analyses can be carried out within regions of mutual validity.

\section{Synthetic Data Study\label{sec:synthetic}}

To test the calculation of the window observables from lattice QCD data, we generate a synthetic dataset from JAM3D$\ast$~\cite{Gamberg:2022kdb,Cocuzza:2023oam}, shown in Fig.~\ref{fig:data_synthetic}. In the supplementary materials, we show the same analysis on the JAMDiFF dataset~\cite{Cocuzza:2023vqs,Cocuzza:2023oam}. We study the $u-d$ transversity ITD at $\mu=2{\rm \,GeV}$, discretized with a unit spacing using $\nu_{\rm max}=4$. At this $\nu_{\rm max}$, the ITD has not converged to zero, but the extraction can be performed provided extrapolation errors are correctly quantified. In the Supplementary Materials, we show the results with $\nu_{\rm max}=10$. 

Our numerical analyses utilize the CP-projected PDFs, bounded in $x\in[0,1]$,
\begin{equation}
q_\pm(x) = f(x) \pm - f(-x) \,.
\end{equation}
This decomposition is convenient because the CP even, $q_-$, and odd, $q_+$, distributions generate respectively the real and imaginary components of the ITD.

We first construct a ITD from the JAM3D$\ast$ dataset and then apply Gaussian Process Regression (GPR) to infer the PDF from the synthetic ITD data. The implementation follows Ref~\cite{Dutrieux:2024rem} except that we use zero as the prior's mean, which drives the extrapolation of the PDF as $x\to0$. The results are shown in Fig.~\ref{fig:data_synthetic}. At all $x$ the reconstruction systematically overestimates the error of the input, but most dramatically for $x<0.4$. This may give the misleading impression that any information taken from low $x$ regimes will be similarly inaccurate.

\begin{figure}
\centering
    \includegraphics[width=\linewidth]{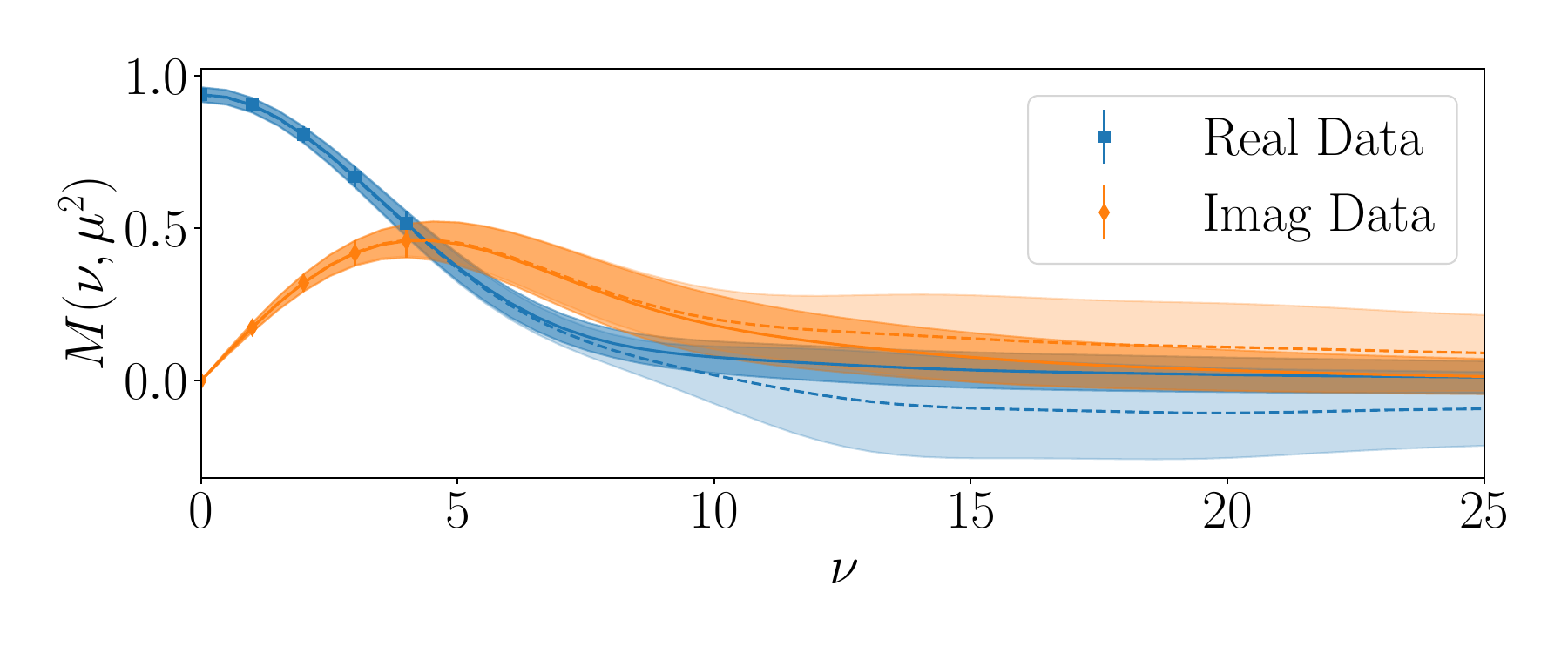}
    \includegraphics[width=\linewidth]{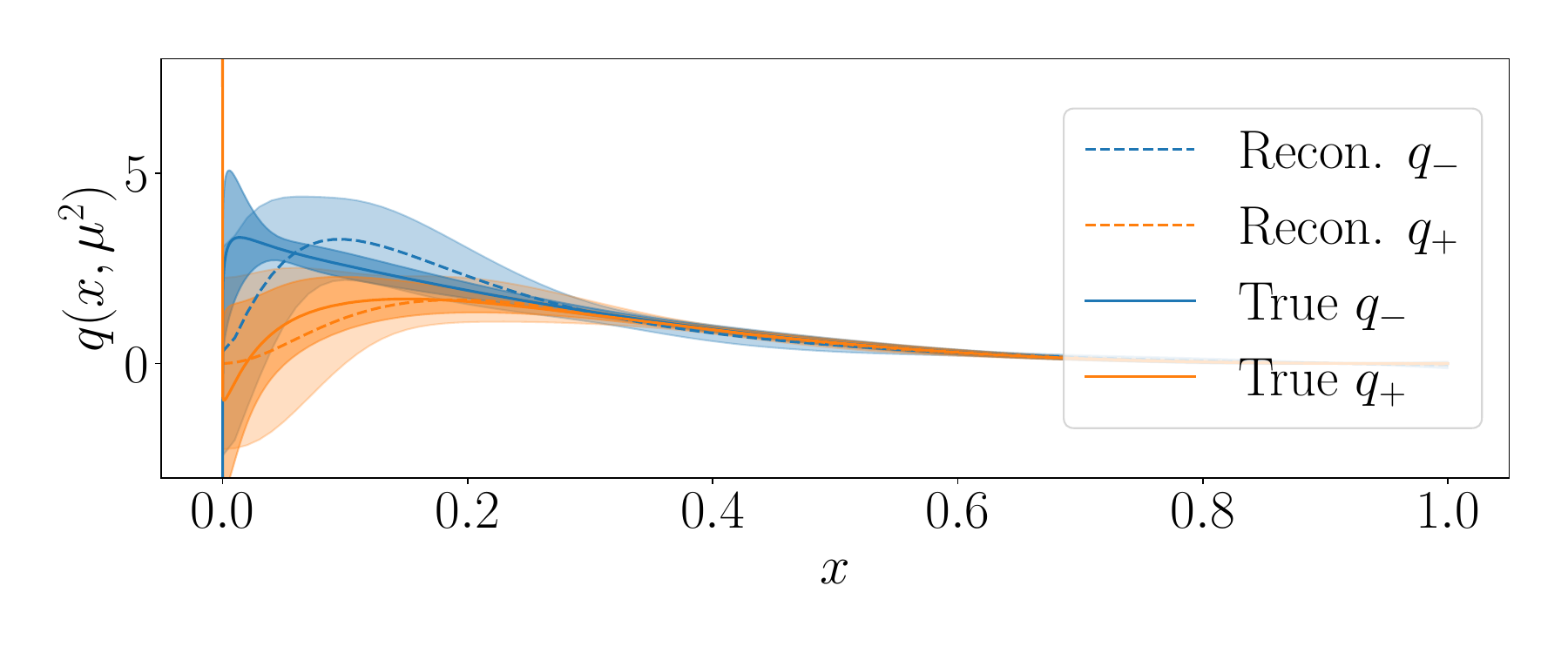}
    \caption{The GPR of synthetic data from JAM3D$\ast$. The GPR reconstruction (dashed line, lighter band) for the $CP$ even (blue) and $CP$ odd (orange) PDFs compared to the original input (solid line, darker band) for (Upper) the ITD and (Lower) the PDF.}\label{fig:data_synthetic}
\end{figure}

For our example of the window observables, we choose $x_-=0.1$ within the modern range of Bjorken-$x$ in experimental datasets for the transversity distribution. With $x_->0$ we can avoid biases at asymptotically low $x$ in the phenomenological extraction and worse lattice extractions. Although $x_+=0.5$ lies above experimentally-accessible Bjorken-$x$, this choice includes a region of $x$ that is still informed by experimental data, due to the convolutions for evolution and matching to cross sections.  We expect that this $x_+$, which is relatively close to the dataset, avoids the worst extrapolation biases that $x_+\to1$ could exhibit. We study alternative windows in the Supplementary Material.

The reconstructed window observables are compared with the original data in Fig.~\ref{fig:windows_synthetic_pheno}. The relative error of the reconstructions is given in the Supplementary Material. Despite the restricted Ioffe time range, the window observables are remarkably accurate, particularly at small $n$, because the low $\nu$ region is most sensitive to the wider features in $x$. The increased error of the GPR with respect to the input and the deviation in central value are linked to the extrapolation error, which reduces when $\nu_{\rm max}$ is increased.

\begin{figure}
    \centering
    \includegraphics[width=\linewidth]{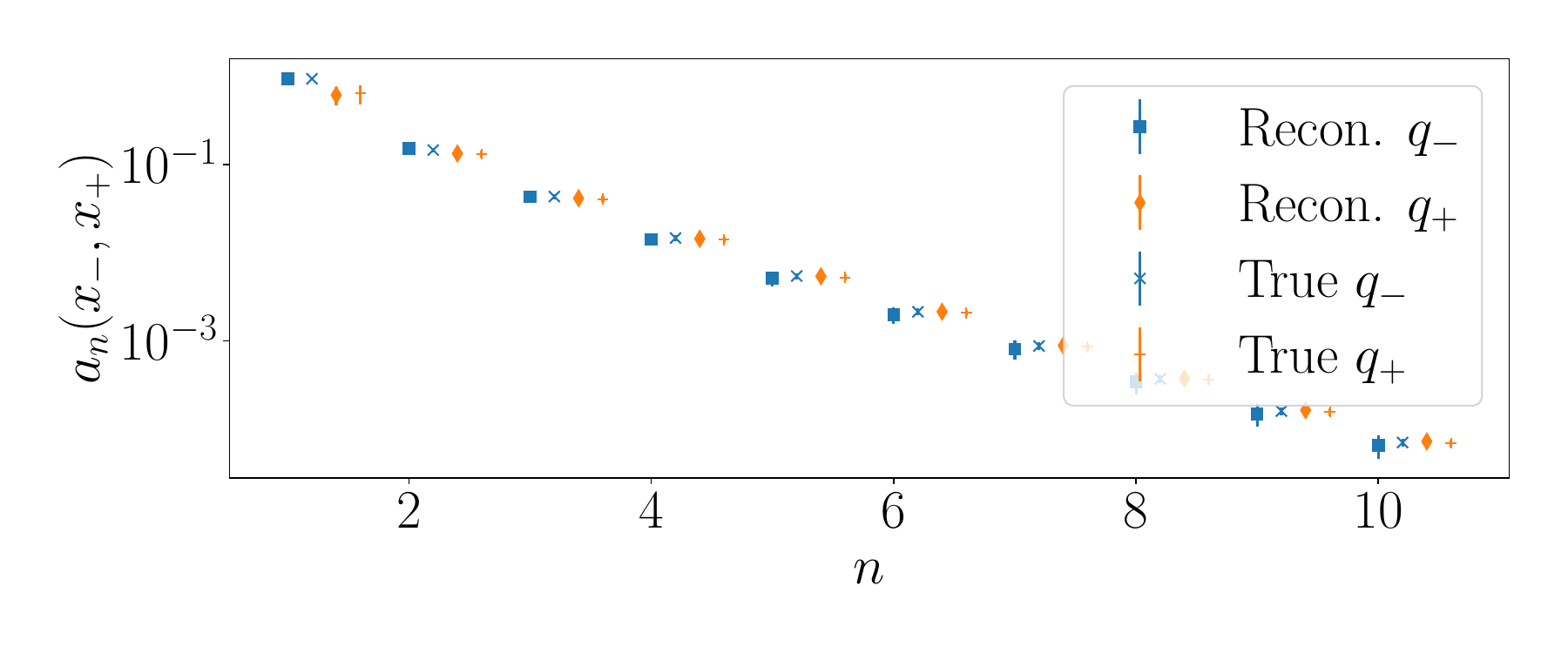}
    \includegraphics[width=\linewidth]{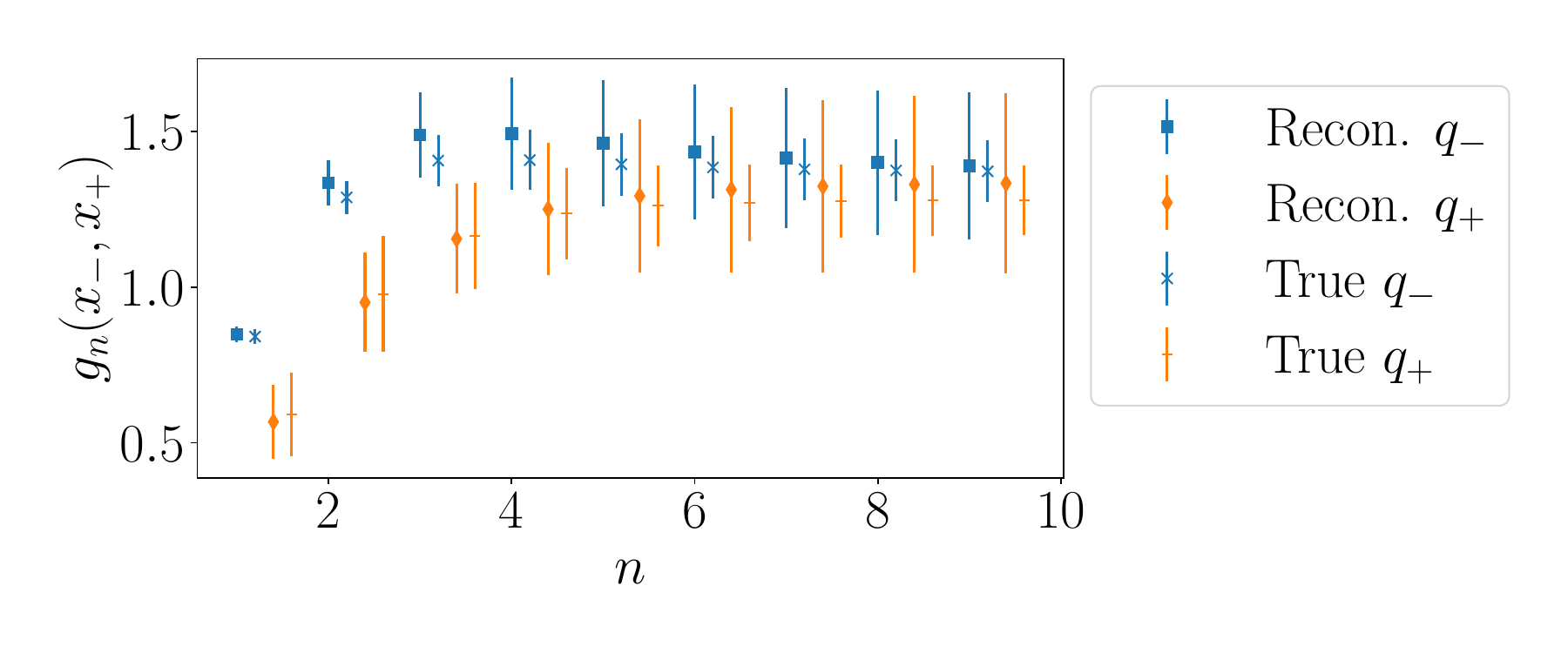}
    \caption{The window moment (upper) and Gaussian window (lower) with $x_-=0.1$ and $x_+=0.5$. }\label{fig:windows_synthetic_pheno}
\end{figure}

The increased uncertainty of the GPR compared to the input distribution point-by-point in $x$ does not translate to a similarly sized increase in the errors of the window observables on the lowest $n$'s. The non-trivial covariance in $x$ from the GPR is visually hidden in Fig.~\ref{fig:data_synthetic}, but properly accounted for in the integrations of window observables. The correlations between data points incorporate the high precision of low Fourier modes, which impacts the precision of slowing varying features that characterize integrals over a wide region of $x$.

Therefore, while the overall $x$-reconstruction of the lattice data only allows a broad qualitative statement of agreement due to its overwhelming uncertainty, the window observables with low $n$ provide a much more stringent benchmarking opportunity between lattice data and phenomenological extractions.

\section{Lattice QCD Data Study}

\begin{figure}
    \centering
    \includegraphics[width=\linewidth]{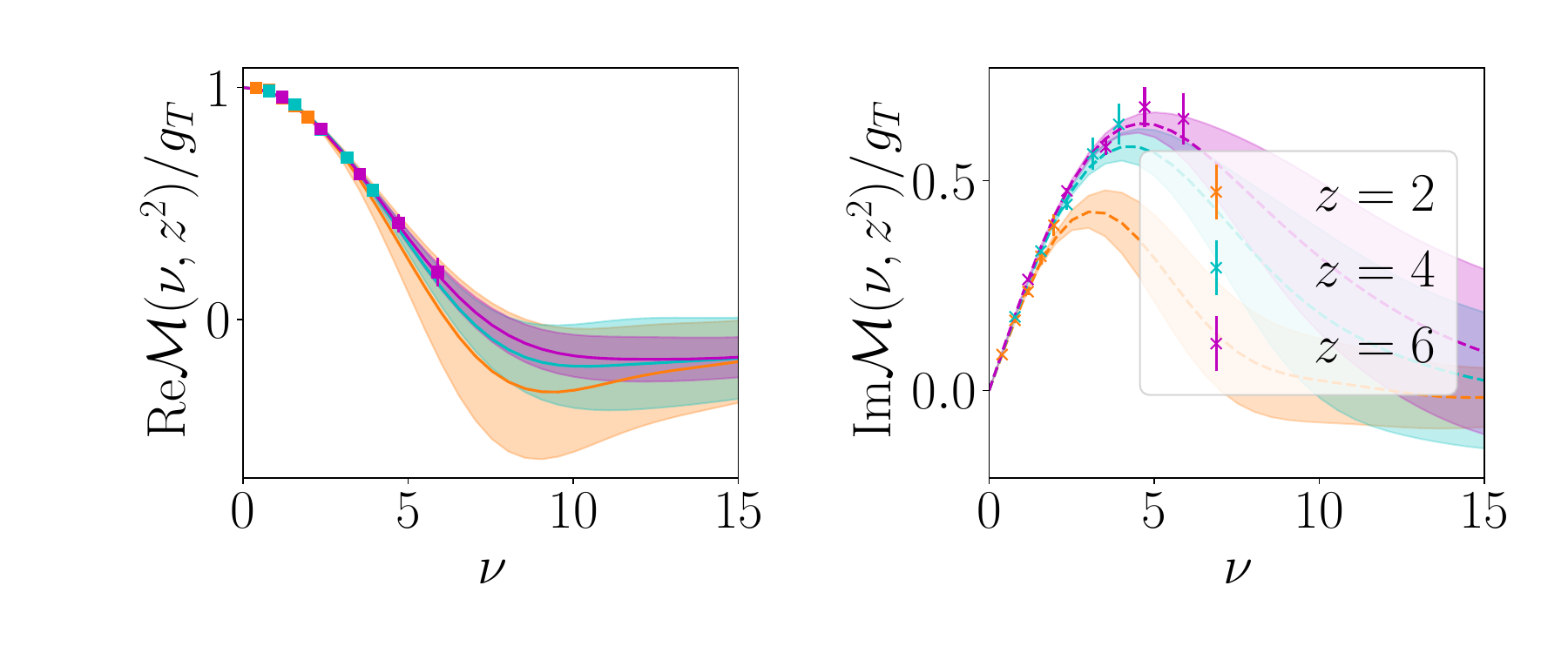}
    \includegraphics[width=\linewidth]{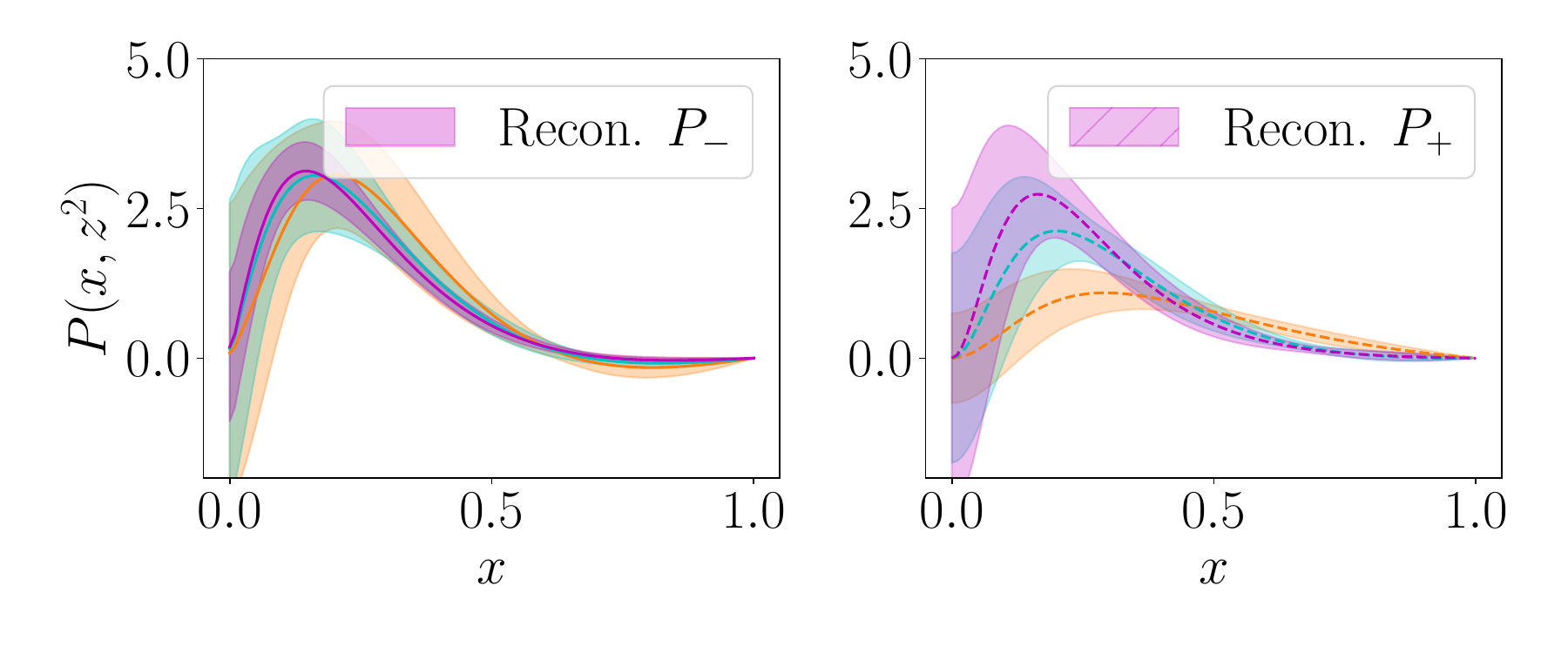}
    \includegraphics[width=\linewidth]{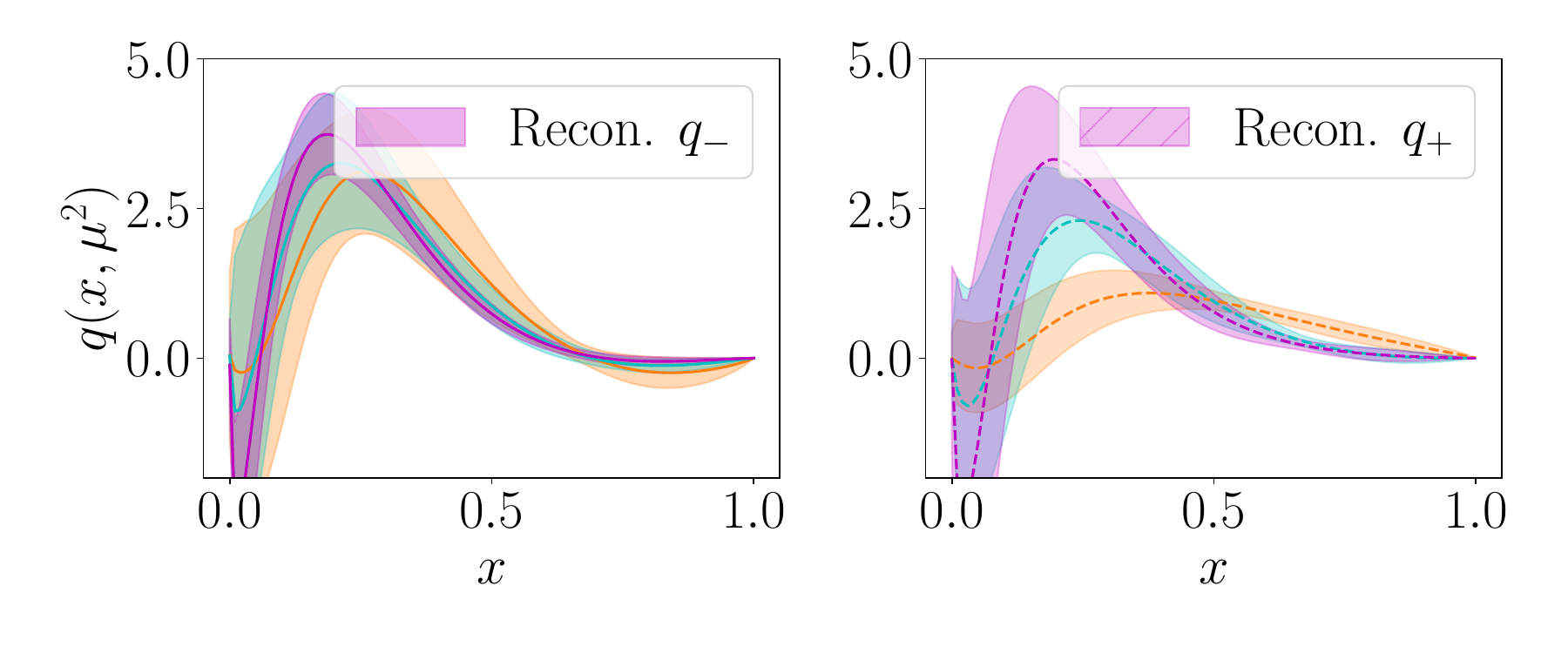}
    \caption{(Upper) The real (squares) and imaginary (crosses) components of the lattice QCD data fit to a GPR. (Middle) The $CP$ even (left) and $CP$ odd (right) pseudo-PDF reconstructions. (Lower) The PDF at $\mu=2$ GeV resulting from LO matching and LL evolution.}\label{fig:data_real}
\end{figure}

We study window observables using lattice QCD data for the renormalization-group invariant (RGI) ratio of pseudo-ITDs~\cite {Karpie:2018zaz} on a single ensemble with lattice spacing $a=0.094$fm and pion mass $m_\pi=358$MeV, described in detail in~\cite{HadStruc:2021qdf}. In our analysis of lattice QCD data, we first obtain the CP-projected pseudo-PDFs $P_\pm(x,z^2)$ from lattice data of the space-like hadronic matrix element using GPR and subsequently match to the PDF. By reducing correlated fluctuations and giving a fixed point at $\nu=0$, the use of RGI ratios helps constrain the GPR substantially but, as a consequence, the lattice data differ from the synthetic data study of the previous section by the normalization, $g_T$. Using the one loop expression~\cite{HadStruc:2021qdf}, the leading-order (LO) matching 
and leading-log (LL) evolution are applied to the pseudo-PDF result to obtain the normalized $\overline{\rm MS}$ PDFs at $\mu=2$ GeV, shown in Fig.~\ref{fig:data_real} for three $z$ values. Note this nomenclature, which conforms to that typically used in global analyses, is uncommon for lattice QCD studies, which typically denote these NLO and NLO + RGR respectively. To help control the evolution and matching of the most poorly constrained models for the $x>0.8$ region, an additional GP prior is added to force the pseudo-PDF to vanish at $x=1$, with little effect on the low $x$ window observables. It is clear that the separations suffer from the limited extent in $\nu$. In the imaginary component the data rises until a peak around $\nu=5$ and falls off again, but short $z$, which fail to reach this peak, tend to underestimate its size. Given available momenta, $z=4a$ is the shortest separation to reach this value. Similar extrapolation biases can more subtly affect the real component. 

\begin{figure}
    \centering
    \includegraphics[width=\linewidth]{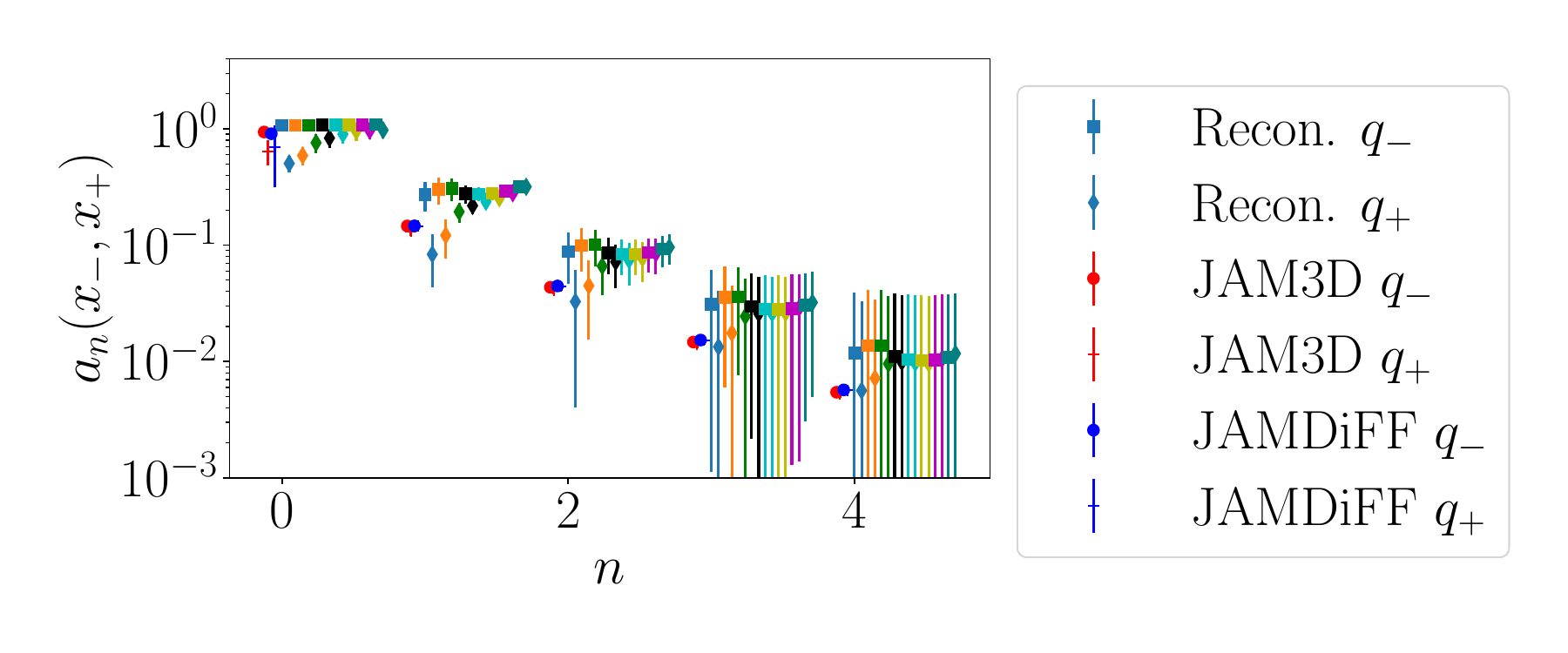}
    \includegraphics[width=\linewidth]{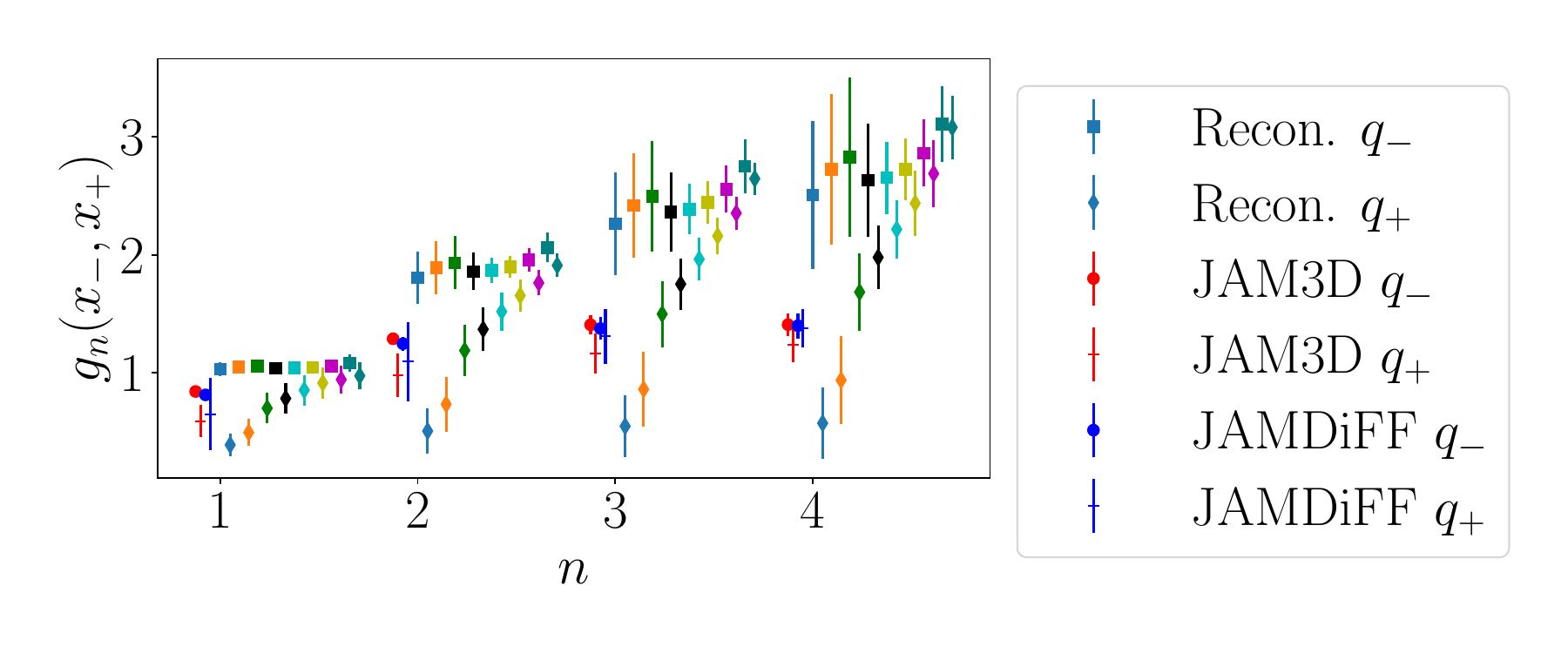}
    \caption{The window moment (upper) and Gaussian window (lower) with $x_-=0.1$ and $x_+=0.5$. The lattice QCD data for $q_-$ (squares) and $q_+$ (diamonds) are given with increasing $z/a$ in the range $[1,8]$ from left to right.}
    \label{fig:windows_real_pheno}
\end{figure}

For comparison to the phenomenological results, the window integrals are multiplied by the $g_T$ normalization obtained from the bare matrix element~\cite{Egerer:2018xgu}, RI-sMom~\cite{Martinelli:1994ty,Sturm:2009kb} renormalization constant~\cite{Park:2021ypf}, and NNLO matching to $\overline{\rm MS}$ scheme~\cite{Gracey:2011fb,Yoon:2016jzj}. We show the window observables in Fig~\ref{fig:windows_real_pheno}. They are generally of the same size with an upward trend with $z$, particularly the Gaussian window. The errors also tend to decrease slightly as $z$ increases, because the larger Ioffe-time extent leads to more constrained reconstructions within the window. 

Future study with many ensembles is required to disentangle the effects from discretization, quark mass, higher twist, higher order perturbation theory, and GP reconstruction error which all contribute to this residual $z$ dependence and cause discrepancies with phenomenology. These limitations are precisely what hinders point-by-point comparison in $x$ but, as demonstrated in the previous section, window observables have smaller reconstruction errors, which could otherwise hide discrepancies. Despite the limited control of lattice QCD systematics, the results are quite similar to the study from synthetic data. At larger $n$ the lattice results begin to lose signal as the observable becomes more akin to a point-like reconstruction. The values at the smallest $n$ have uncertainties comparable to phenomenology, demonstrating that precision benchmarking will become a possibility as lattice QCD results improve.

\section{Conclusion and Outlook}\label{sec:conc}
Hadrons are dynamic, strongly-coupled systems composed of quarks and gluons. Asymptotic freedom ensures that the low-energy structure of hadrons can be factorized into universal, nonperturbative parton distribution functions. 

A fundamental limitation of global analyses of parton distributions is the limited range of Bjorken-$x$ available from experimental data. Similarly, a fundamental limitation of lattice QCD calculations is the limited range of noisy data.  
To benchmark lattice QCD calculations and to compare different approaches with phenomenological extractions, it is desirable to find observables with controlled errors. In this work we propose one set of such observables, dubbed window observables. These quantities are sensitive to wide features in a region in $x$, allowing for more precise comparisons between lattice QCD results than point-by-point comparisons, while constrained to kinematic regions where global analyses are most reliable.

We demonstrate the extraction of the window observables from synthetic and real lattice data for the quark transversity of the nucleon, generated for a restricted range of Ioffe times. Lattice QCD results are most precise in low Fourier modes of the PDF, causing wider window observables to be dramatically more precise than suggested by a point-by-point comparison in $x$. 
In future calculations with data at multiple lattice spacings, pion masses, and physical volumes, the window observables will provide a high precision tool for cross-validation of lattice QCD and phenomenology.

\section{Acknowledgments}

We are grateful to Herv{\'e} Dutrieux for extensive discussions and helpful comments on the manuscript. This project was supported by the U.S.~Department of Energy, Office of Science, Contract No.~DE-AC05-06OR23177, under which Jefferson Science Associates, LLC operates Jefferson Lab. KO was supported in part by U.S.~DOE Grant \mbox{\#DE-FG02-04ER41302} and would like to acknowledge the hospitality of the American Academy in Rome, where he spent part of his sabbatical. This research was funded, in part (SZ), by l’Agence Nationale de la Recherche (ANR), project ANR-23-CE31-0019. CJM is supported in part by U.S.~DOE Grant \mbox{\#DE-SC0025908} and in part by U.S.~DOE ECA \mbox{\#DE-SC0023047}. This work has benefited from the collaboration enabled by the Quark-Gluon Tomography (QGT) Topical Collaboration, U.S.~DOE Award \mbox{\#DE-SC0023646}. Computations for this work were carried out in part on facilities of the USQCD Collaboration, which are funded by the Office of Science of the U.S.~Department of Energy. The authors acknowledge William \& Mary Research Computing for providing computational resources and/or technical support, which were provided by contributions from the National Science Foundation (MRI grant PHY-1626177), and the Commonwealth of Virginia Equipment Trust Fund. In addition, this work used resources at NERSC, a DOE Office of Science User Facility supported by the Office of Science of the U.S. Department of Energy under Contract \#DE-AC02-05CH11231, as well as resources of the Oak Ridge Leadership Computing Facility at the Oak Ridge National Laboratory, which is supported by the Office of Science of the U.S. Department of Energy under Contract No. \mbox{\#DE-AC05-00OR22725}. 
The authors acknowledge support as well as computing and storage resources by GENCI on Adastra (CINES), Jean-Zay (IDRIS) under project (2020-2024)-A0080511504.
The software codes {\tt Chroma} \cite{Edwards:2004sx}, {\tt QUDA} \cite{Clark:2009wm, Babich:2010mu}, {\tt QPhiX} \cite{QPhiX2}, and {\tt Redstar} \cite{Chen:2023zyy} were used in our work. The authors acknowledge support from the U.S. Department of Energy, Office of Science, Office of Advanced Scientific Computing Research and Office of Nuclear Physics, Scientific Discovery through Advanced Computing (SciDAC) program, and of the U.S. Department of Energy Exascale Computing Project (ECP). The authors also acknowledge the Texas Advanced Computing Center (TACC) at The University of Texas at Austin for providing HPC resources, like Frontera computing system~\cite{frontera} that has contributed to the research results reported within this paper. 

\FloatBarrier
\appendix 

\section{Supplementary Material}

\section{Window observables as Ioffe time Integrals}
The window observables can be written as integrals in Ioffe time. Let us start from the general convolution without bounds on $x$
\begin{equation}
    W_n(x_\pm) = \int_{-\infty}^\infty dx\,w_n(x; x_\pm) q(x)
\end{equation}
following the assumption that $q(|x|>1)=0$. The convolution theorem leads us to the analogous Fourier integration
\begin{equation}
    W_n(x_\pm) = \int_{-\infty}^\infty \frac{d\nu}{2\pi} \tilde{w}_n(\nu; x_\pm) M(-\nu)
\end{equation}
where $\tilde{w}(\nu)$ is the inverse FT of $w(x)$.

The Gaussian window given in Eq.~\eqref{eq:gauss_window_def}, which limits the range of $x$ to the vicinity of the midpoint between $x_\pm$, has an inverse FT
\begin{equation}
    \tilde{w}_n(\nu) = \exp\bigg[i\nu \frac{x_++x_-}2- \frac{(x_+-x_-)^2 \nu^2}{2n^2} \bigg] \,.
\end{equation}
where $n$ controls how localized the Gaussian is within the window $[x_-,x_+]$. 

The window moment can be written using step functions $\theta$
\begin{equation}
    a_n(x_-,x_+) = \int_{-\infty}^\infty dx\, (\theta(x-x_-) - \theta(x-x_+)) x^{n-1} q(x)\,.
\end{equation}
For the integration bounds, there is an assumption that the PDF vanished outside the bounds of $x\in [-1,1]$, or $x\in [0,1]$ when $CP$ projected, but with the limited bounds $|x_\pm|<1$ used in practice this assumption does not matter.
The step functions can be replaced by approximations such as a sigmoid or arctangent function, with limited modifications to the following derivation, but in the following we use Heaviside step functions. 

Inserting the inverse Fourier transforms of $q$ and $\theta$, we obtain
\begin{eqnarray}
a_n(x_-,x_+) =   \int \frac{d\nu }{2\pi} (e^{i x_-\nu} - e^{i x_+\nu}) \widetilde{\theta}(\nu) \nonumber\\  \int \frac{d\nu'}{2\pi}  M(\nu') \int dx x^{n-1} e^{-i x (\nu+\nu')} 
\end{eqnarray}
where $\widetilde{\theta}(\nu) $ is the inverse Fourier transform of $\theta(x)$, 
\begin{equation}
\widetilde\theta(\nu)=\pi\delta(\nu) - \frac{i}{\nu}\,.
\end{equation}
The $x$ integrals can be evaluated as $\delta(\nu+\nu')$ or its derivatives
\begin{eqnarray}
a_n(x_-,x_+) =   \int \frac{d\nu d\nu'}{(2\pi)} (e^{i x_-\nu} - e^{i x_+\nu}) \widetilde{\theta}(\nu) \nonumber \\ \times M(\nu') (-i)^{n-1} \delta^{(n-1)}(\nu+\nu')   \,.
\end{eqnarray}

Given that the ITD converges to 0 as $\nu\to \infty$, one can integrate by parts so that
derivatives of delta functions give derivatives of the ITD
\begin{equation}
a_n(x_-,x_+) =    i^{n-1} \int \frac{ d\nu}{(2\pi)} (e^{i x_-\nu} - e^{i x_+\nu}) \widetilde{\theta}(\nu) M^{(n-1)}(-\nu)\,.
\end{equation}
Inserting $\widetilde{\theta}(\nu)$, the delta function term will vanish since the phase terms cancel at $\nu=0$ while the derivative of the ITD remains finite. The other term can be integrated by parts to give
\begin{equation}
    a_n(x_-,x_+) = \frac{i^{n-2}}{4\pi^2} \int d\nu M(-\nu) \frac{\partial^{n-1}}{\partial \nu^{n-1}} \left[ \frac{e^{ix_-\nu}- e^{ix_+\nu}}\nu \right]\,.
\end{equation}
For $n=2$ this becomes
\begin{align}
    a_2(x_-,x_+) & = \frac{1}{2\pi} \int \mathrm{d}\nu \frac{M(-\nu)}{\nu^2} \nonumber \\  
 & \quad \times \Big[ (i x_-\nu-1)e^{ix_-\nu} - (i x_+\nu-1)e^{ix_+\nu} \Big]
\end{align}
The expression for general $n$ is given by 
\begin{align}
    a_n(x_-,x_+) &  = \frac{i^{n-2}}{2\pi} \int d\nu \frac{M(-\nu)}{\nu^n} \nonumber \\
    & \quad \times \Big[ e^{ix_-\nu}p_{n-1}(ix_-\nu) - e^{ix_+\nu}p_{n-1}(ix_+\nu) \Big]
\end{align}
where
\begin{equation}
    p_n(x) = -\sum_{k=0}^n \frac{n!}{k!} (-x)^k
\end{equation}
The integrand in the general case is therefore suppressed by an additional power of $\nu$ compared to the integrands needed to obtain PDFs or pseudo-PDFs,  which already decay with a power law based upon the low-$x$ divergence~\cite{Braun:1994jq}. 
In the limit of $\nu\to0$, the integrand approaches 
\begin{equation}
\frac{(x_-)^n - (x_+)^n}{2\pi n}
\end{equation}
 which vanishes for even $n$ when $x_-=-x_+$. This exact value can be used as a fixed point to guide the interpolation and extrapolation of data required to perform the integral in $\nu$ space.

\section{Relative error of JAM3D$\ast$ reconstructions}
\begin{figure}
    \centering
    \includegraphics[width=\linewidth]{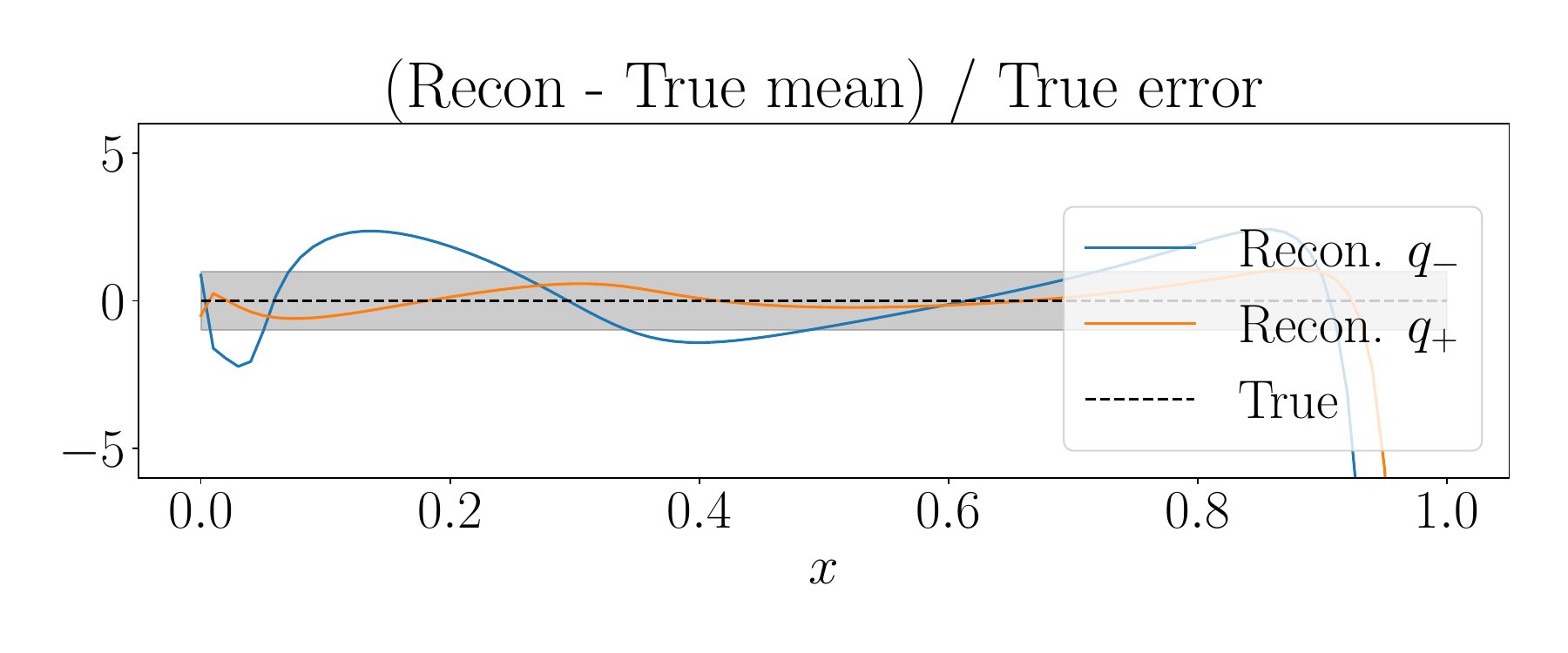}
    \includegraphics[width=\linewidth]{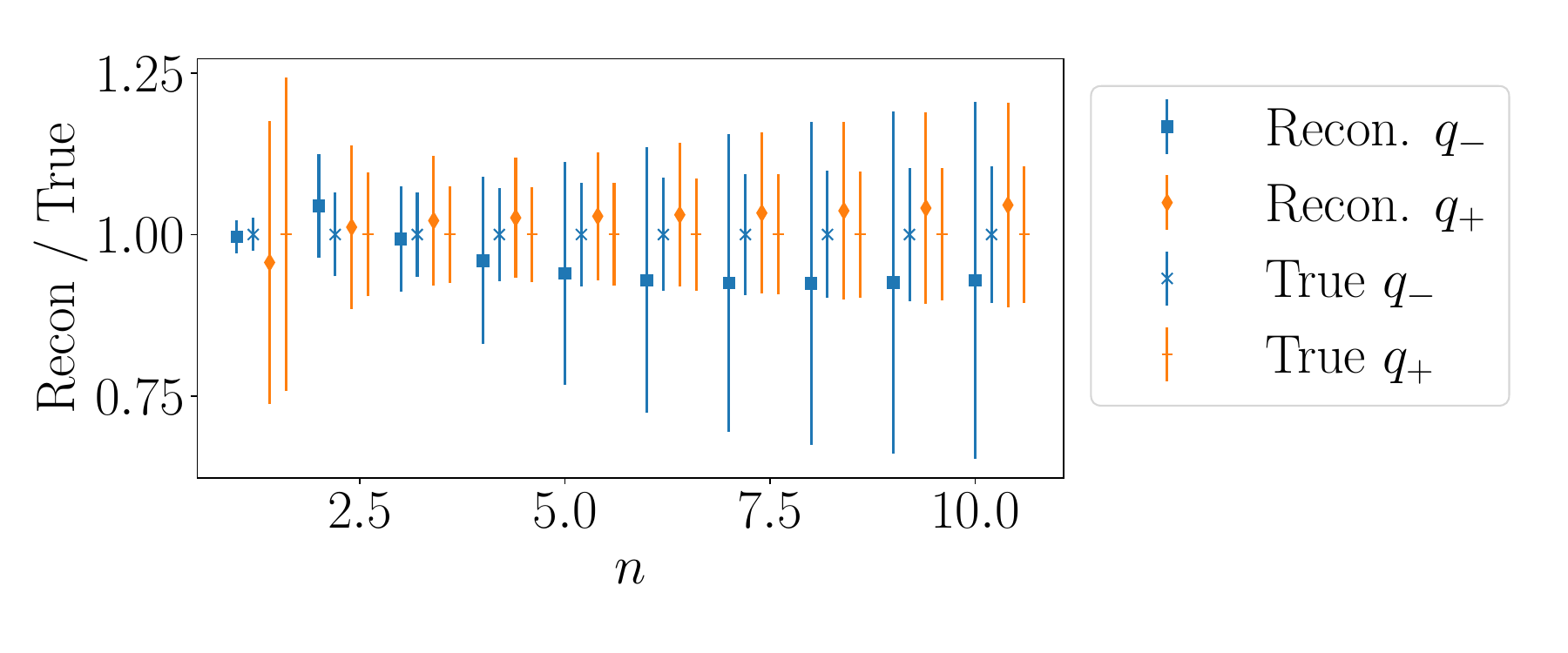}
    \includegraphics[width=\linewidth]{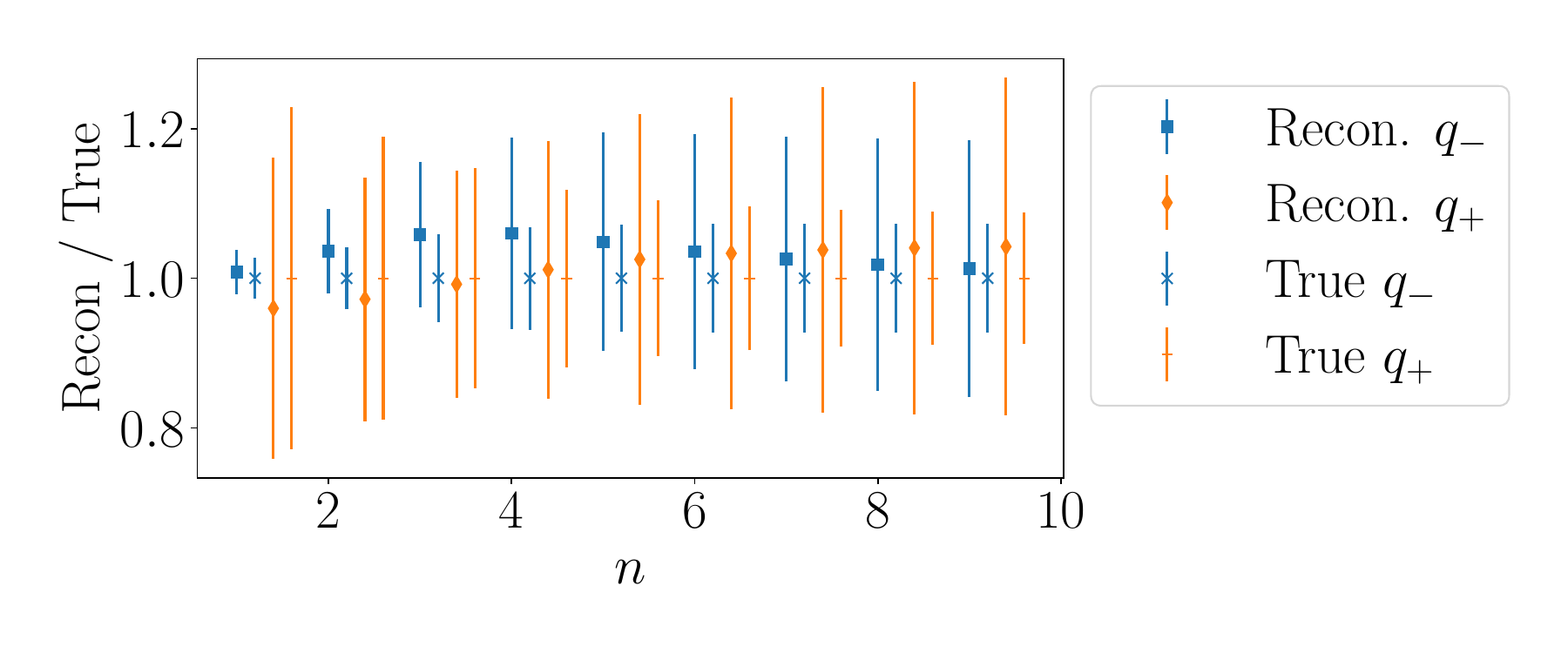}
    \caption{The relative error in the reconstructions of the PDF (upper), Window moments (middle), and Gaussian windows (lower).}\label{fig:relative_synthetic}
\end{figure}

Fig.~\ref{fig:relative_synthetic} shows the relative accuracy of the PDF and window observable reconstructions in Figs.~\ref{fig:data_synthetic} and~\ref{fig:windows_synthetic_pheno}. In the PDF reconstruction, the lower $x$ regime, specifically in the window studied $x\in [0.1,0.5]$, systematically over estimates the error point by point. The lower $n$ window moment integrals on the other hand reconstruct the value with significantly less error than the point by point reconstruction would have implied. The lowest $n$ actually reproduces the original error's size while larger $n$ give the systematic overestimation of point by point reconstruction. 

\section{Results from JAMDiFF}
\begin{figure}
\centering
    \includegraphics[width=\linewidth]{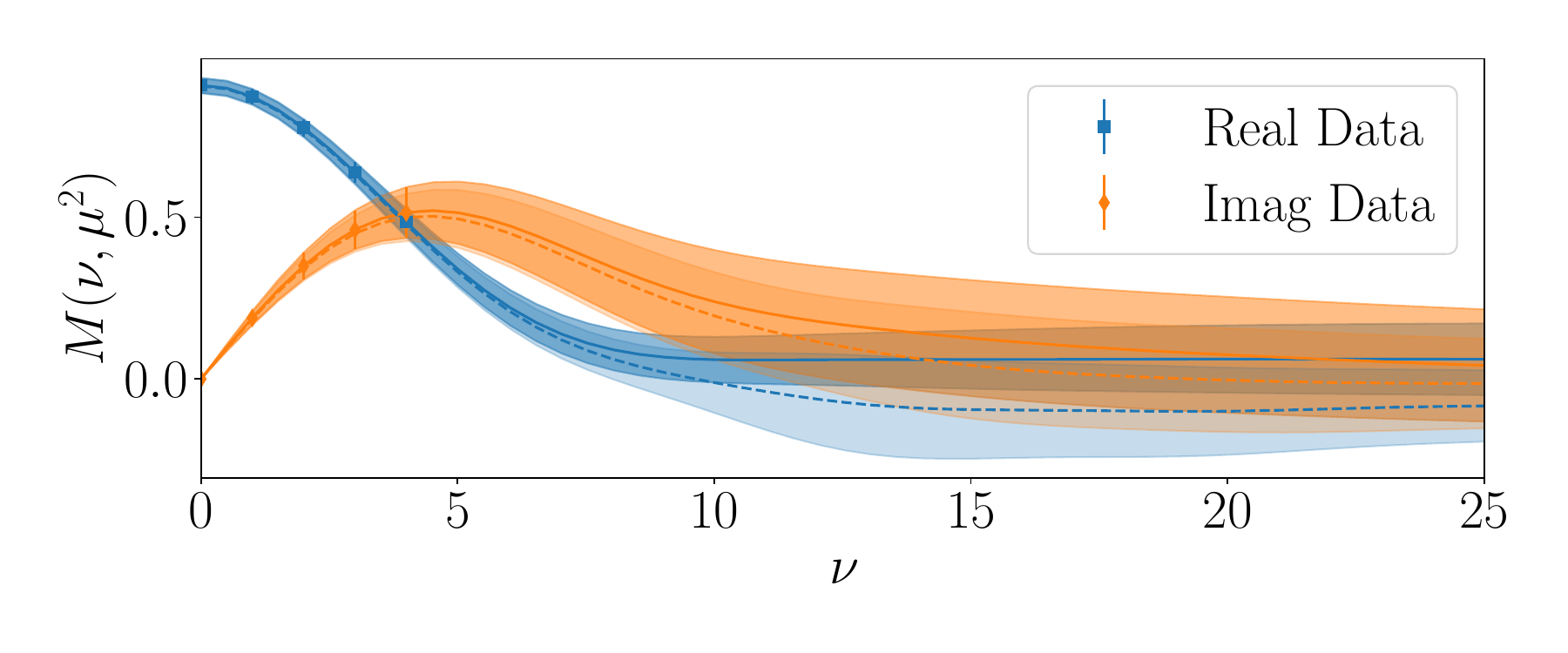}
    \includegraphics[width=\linewidth]{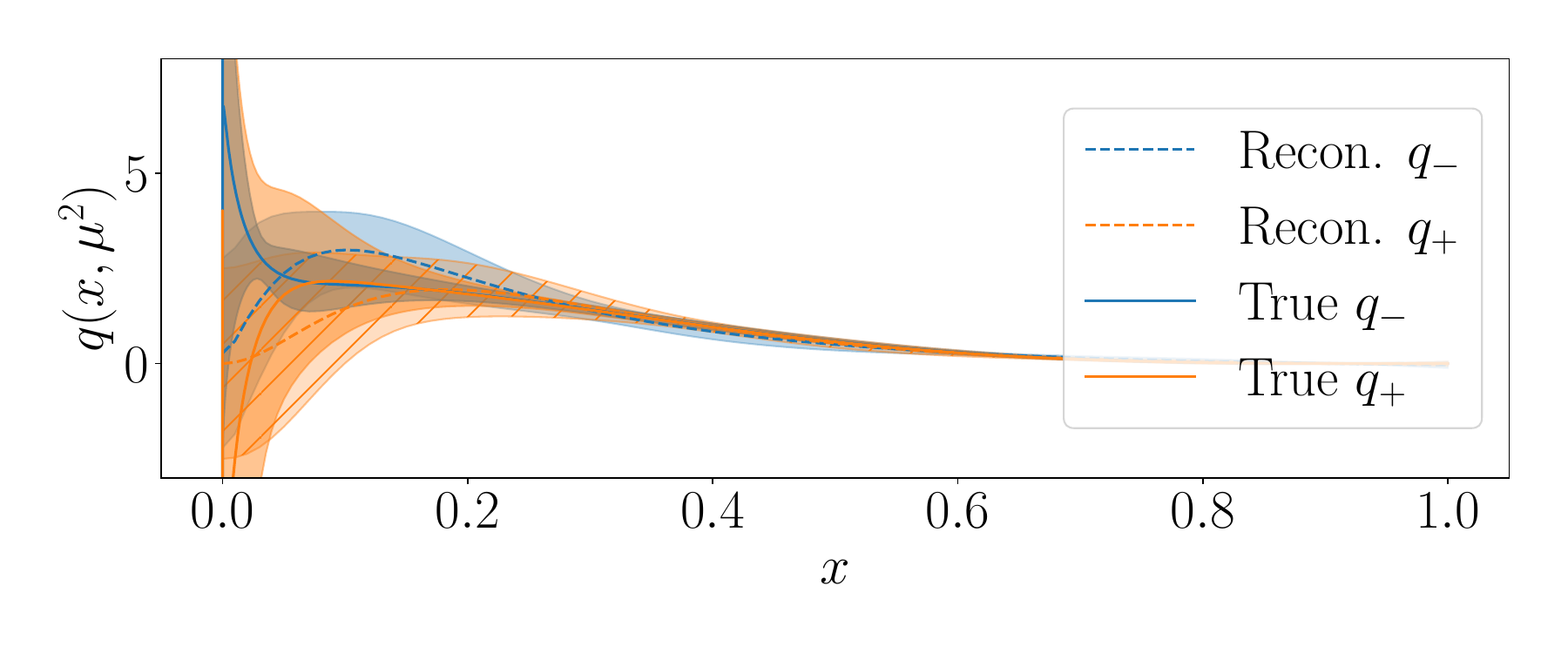}
    \caption{The same as Fig.~\ref{fig:data_synthetic} except with data generated from JAMDiFF results except the reconstructed $q_+$ PDF is hatched for visibility.  }\label{fig:data_synthetic_diff}
\end{figure}

\begin{figure}
    \centering
    \includegraphics[width=\linewidth]{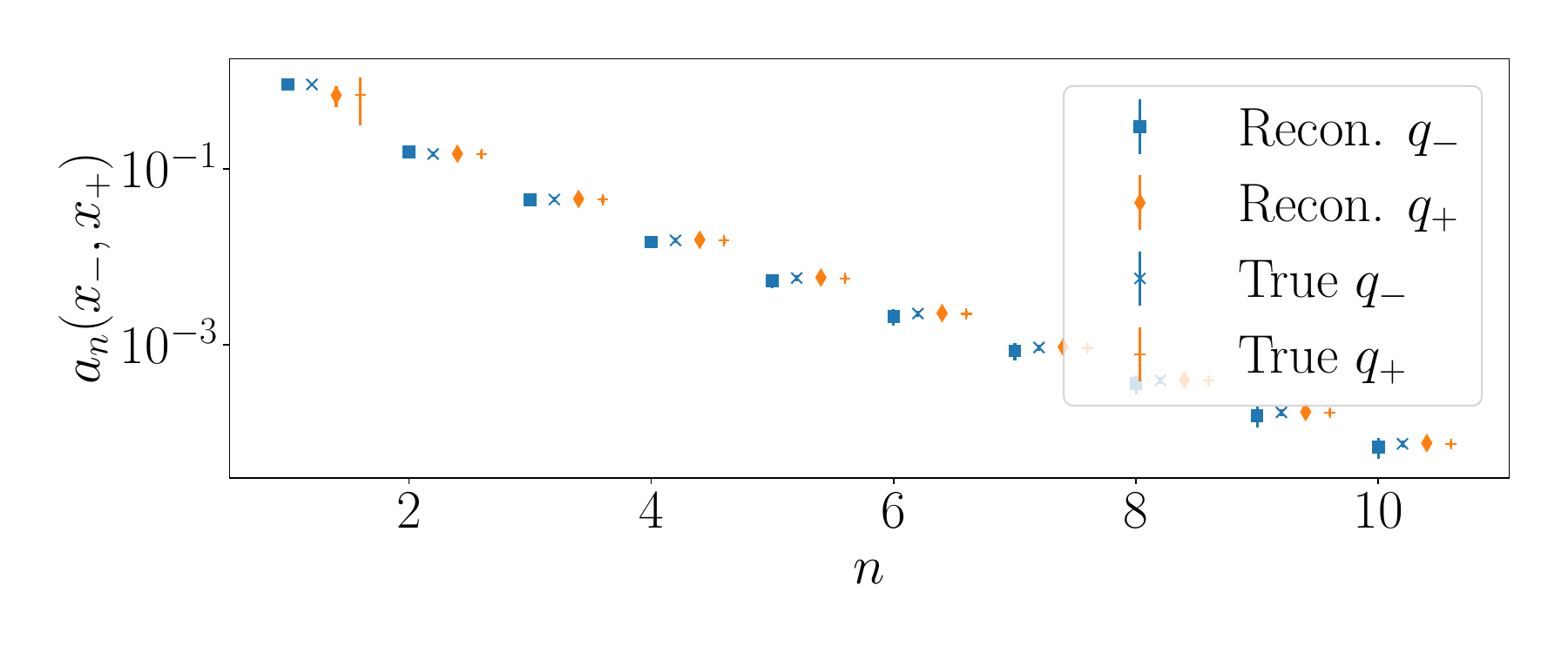}
    \includegraphics[width=\linewidth]{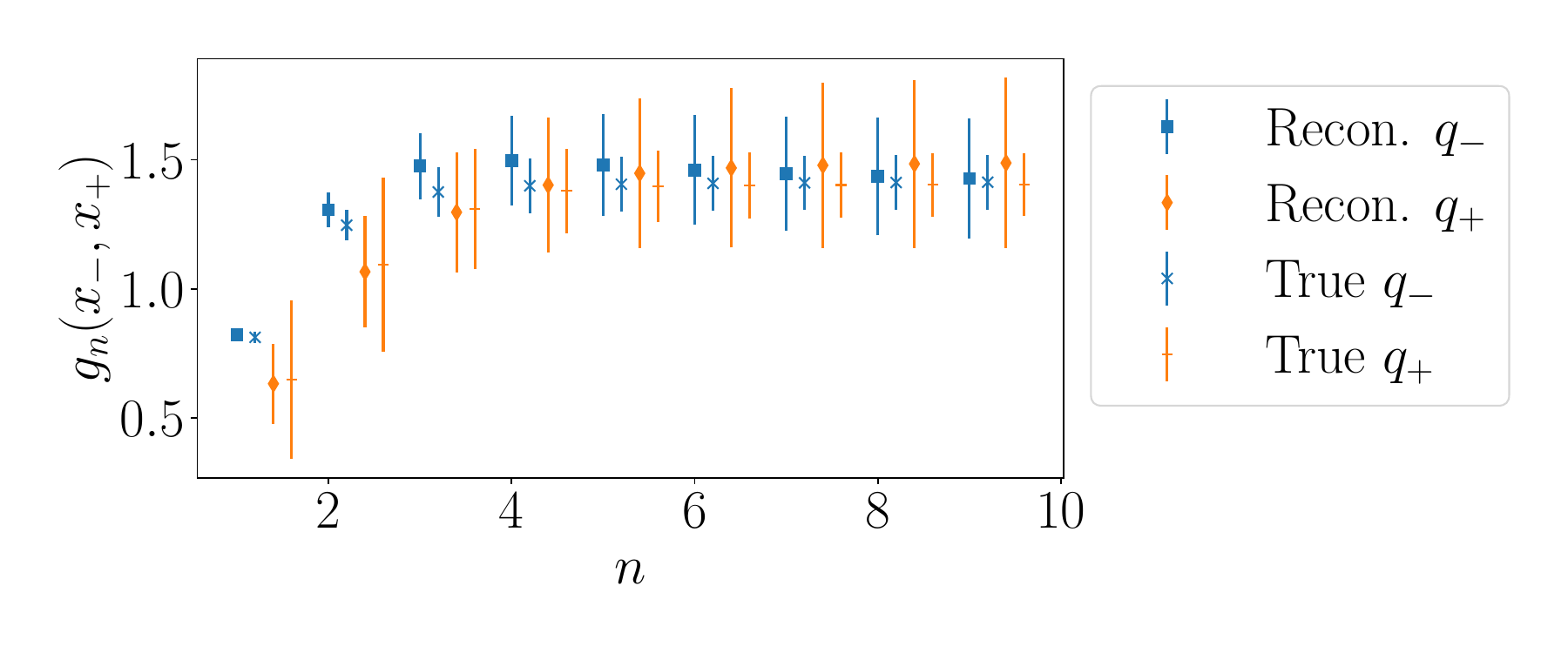}
    \caption{The same as Fig.~\ref{fig:windows_synthetic_pheno} with data generated from JAMDiFF results.}\label{fig:windows_synthetic_pheno_diff}
\end{figure}

Figs.~\ref{fig:data_synthetic_diff} and~\ref{fig:windows_synthetic_pheno_diff} are the same Figs.~\ref{fig:data_synthetic} and~\ref{fig:windows_synthetic_pheno} respectively except the synthetic data were generated with the JAMDiFF results. As constructed, the phenomenological CP odd isovector $u-d$ transversity PDF $q_+$ has larger uncertainties at low $x$ than those from JAM3D$\ast$. The reconstruction of the PDF from the imaginary component data results in an underestimation of this large uncertainty.

\section{Results from JAM3D$\ast$ with $\nu_{\rm max}=10$}
\begin{figure}
\centering
    \includegraphics[width=\linewidth]{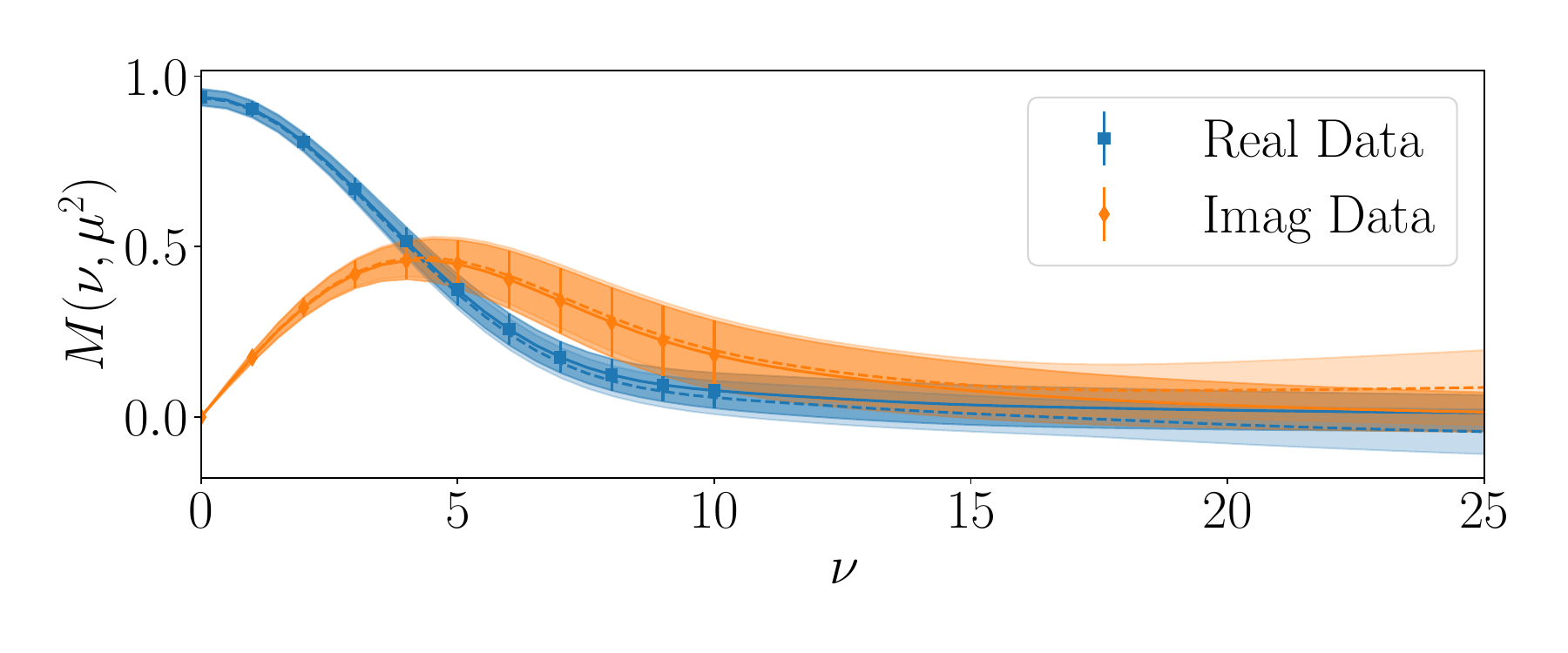}
    \includegraphics[width=\linewidth]{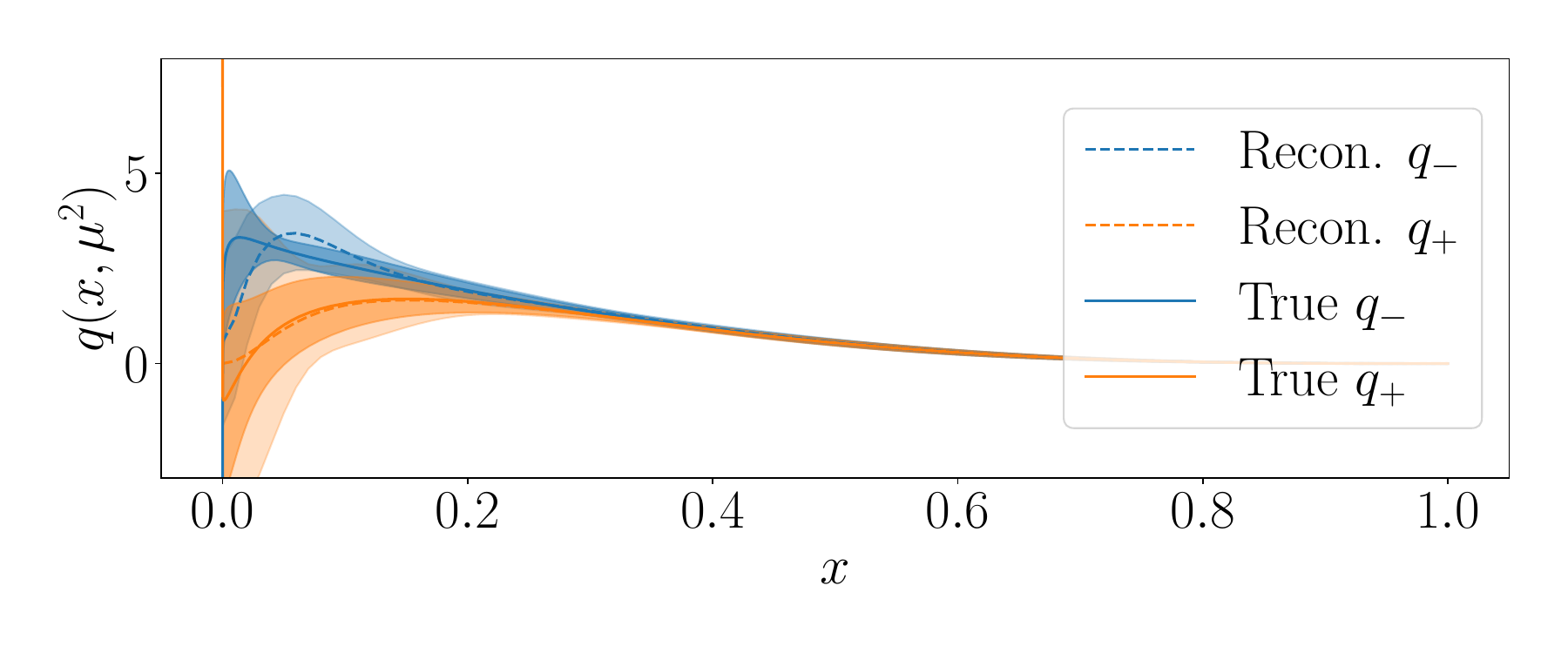}
    \caption{The same as Fig.~\ref{fig:data_synthetic} except with $\nu_{\rm max}=10$.}\label{fig:data_synthetic_10}
\end{figure}

\begin{figure}
    \centering
    \includegraphics[width=\linewidth]{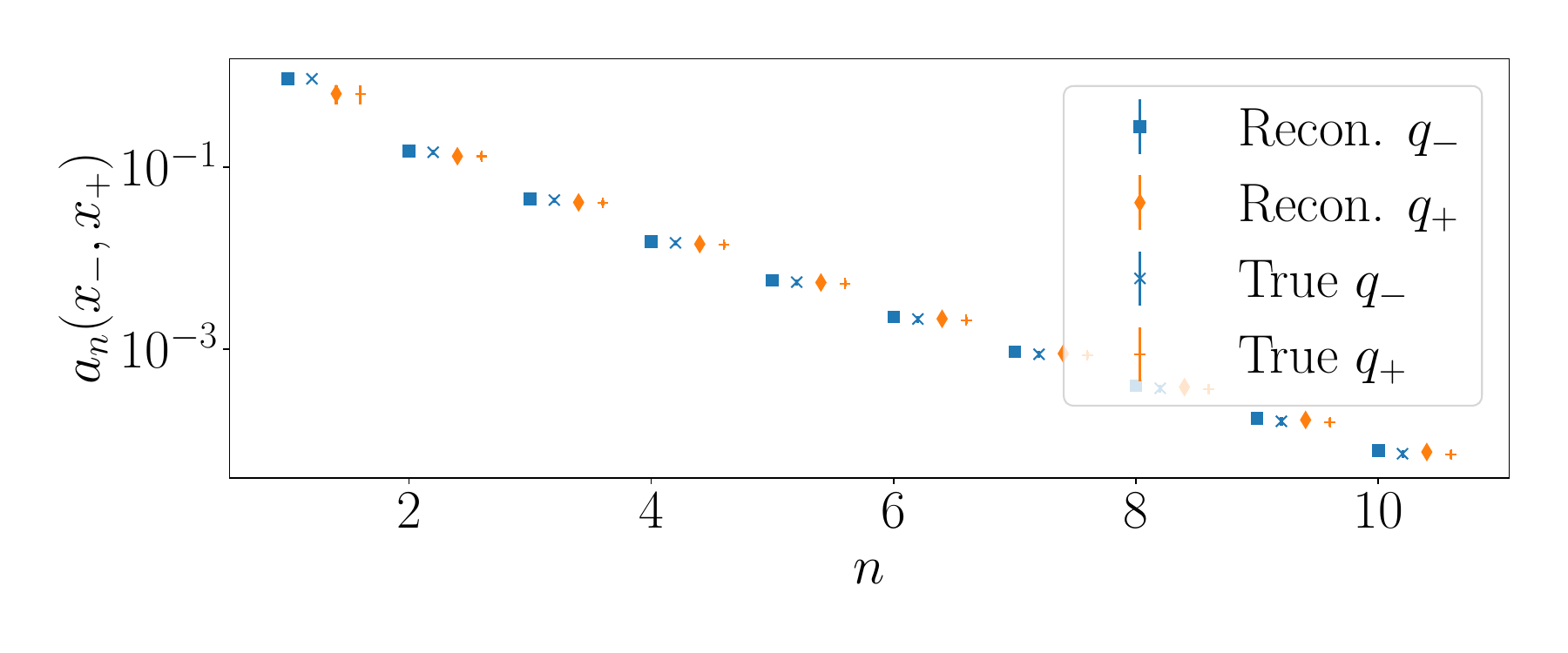}
    \includegraphics[width=\linewidth]{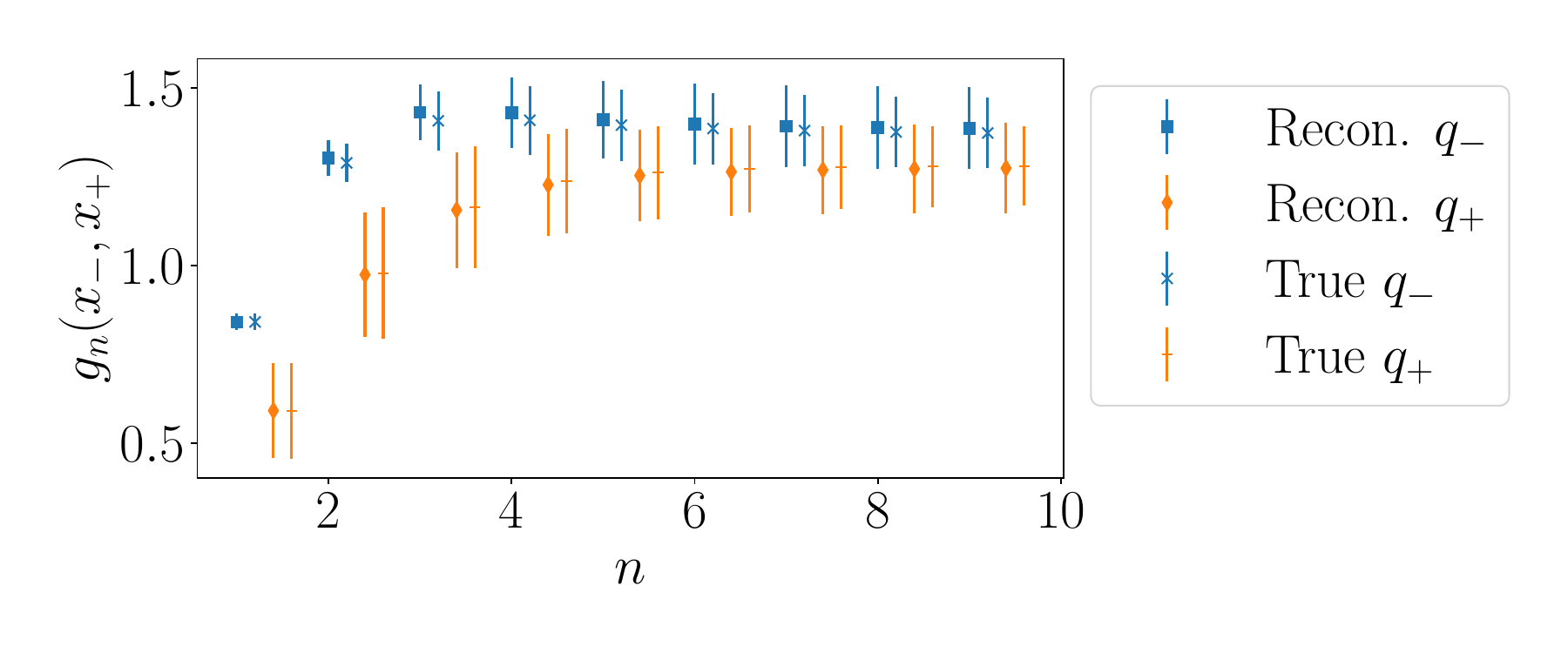}
    \caption{The same as Fig.~\ref{fig:windows_synthetic_pheno} except with $\nu_{\rm max}=10$.}\label{fig:windows_synthetic_pheno_10}
\end{figure}

Figs.~\ref{fig:data_synthetic_10} and~\ref{fig:windows_synthetic_pheno_10} are the same are the same Figs.~\ref{fig:data_synthetic} and~\ref{fig:windows_synthetic_pheno} respectively with a $\nu_{\rm max}=10$. The reconstruction error is significantly more under control. The window observables' errors are more similar to the true value, specifically for larger $n$.

\section{Alternate choices of windows}
\begin{figure}
    \centering
    \includegraphics[width=\linewidth]{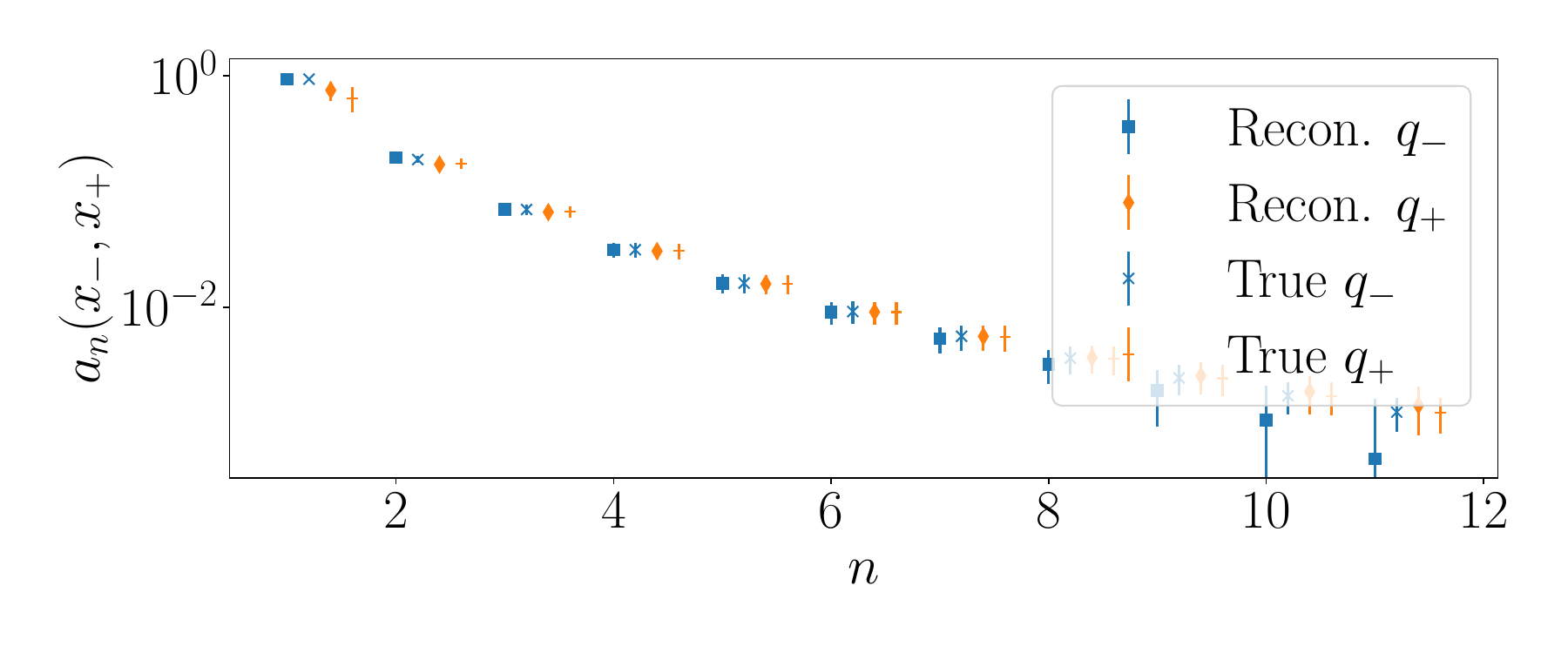}
    \includegraphics[width=\linewidth]{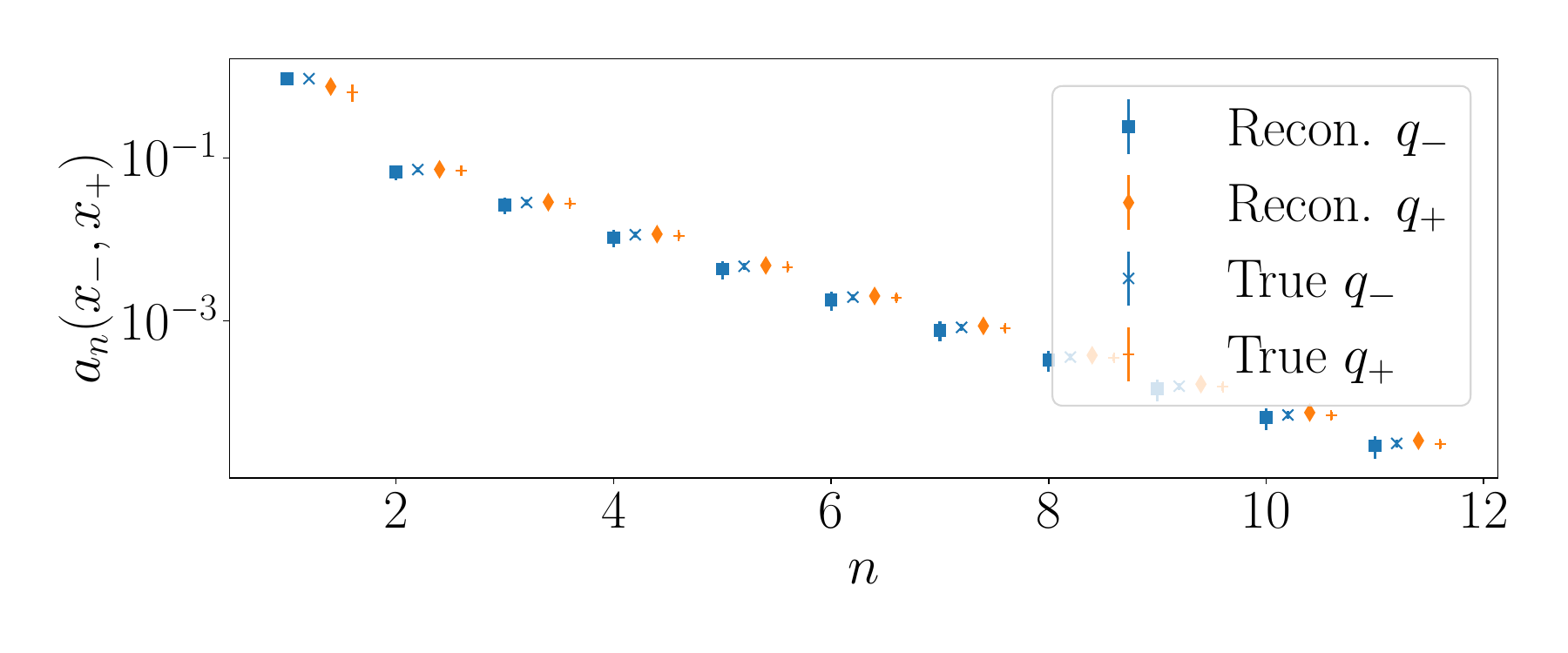}
    \includegraphics[width=\linewidth]{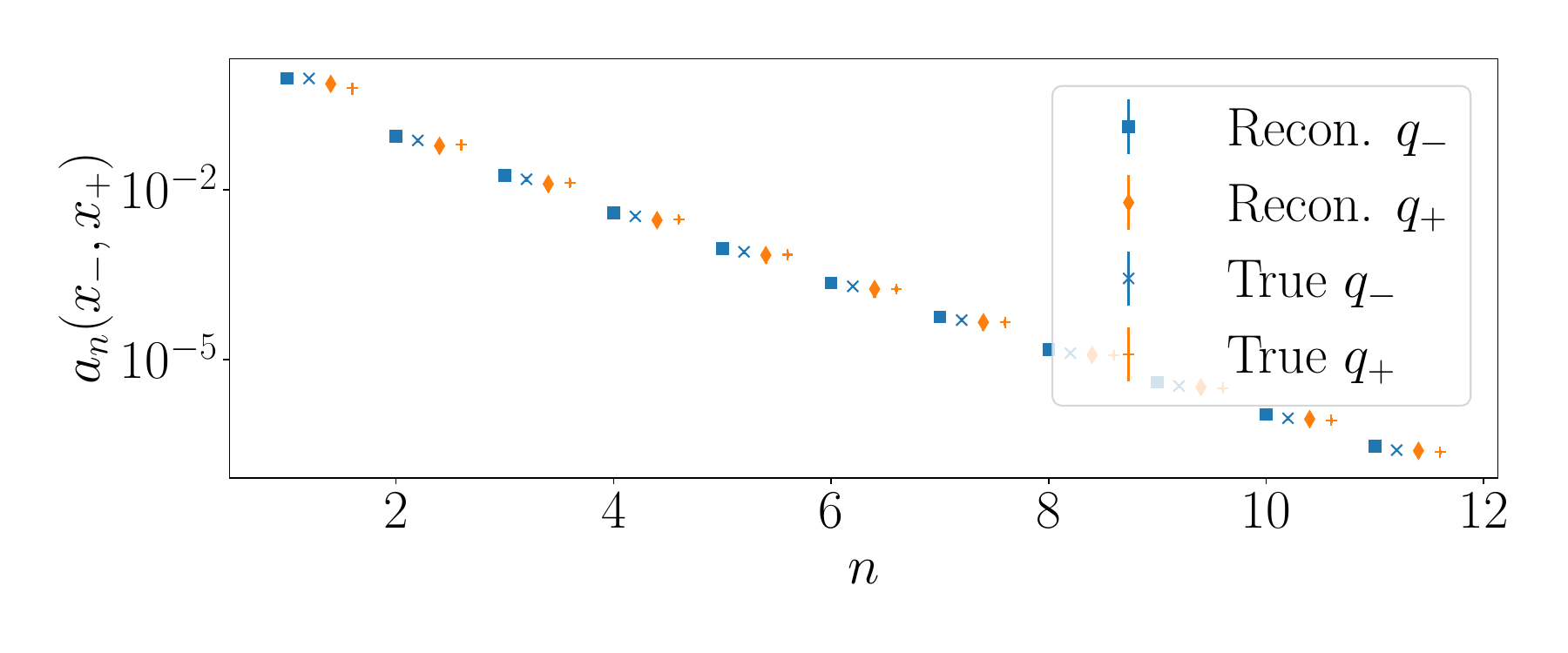}
    \caption{The window moments with, in descending order, $(x_-,x_+)$ = (0.1, 1), (0.3, 0.5), and (0.1, 0.3) . }\label{fig:other_window_moms}
\end{figure}

\begin{figure}
    \includegraphics[width=\linewidth]{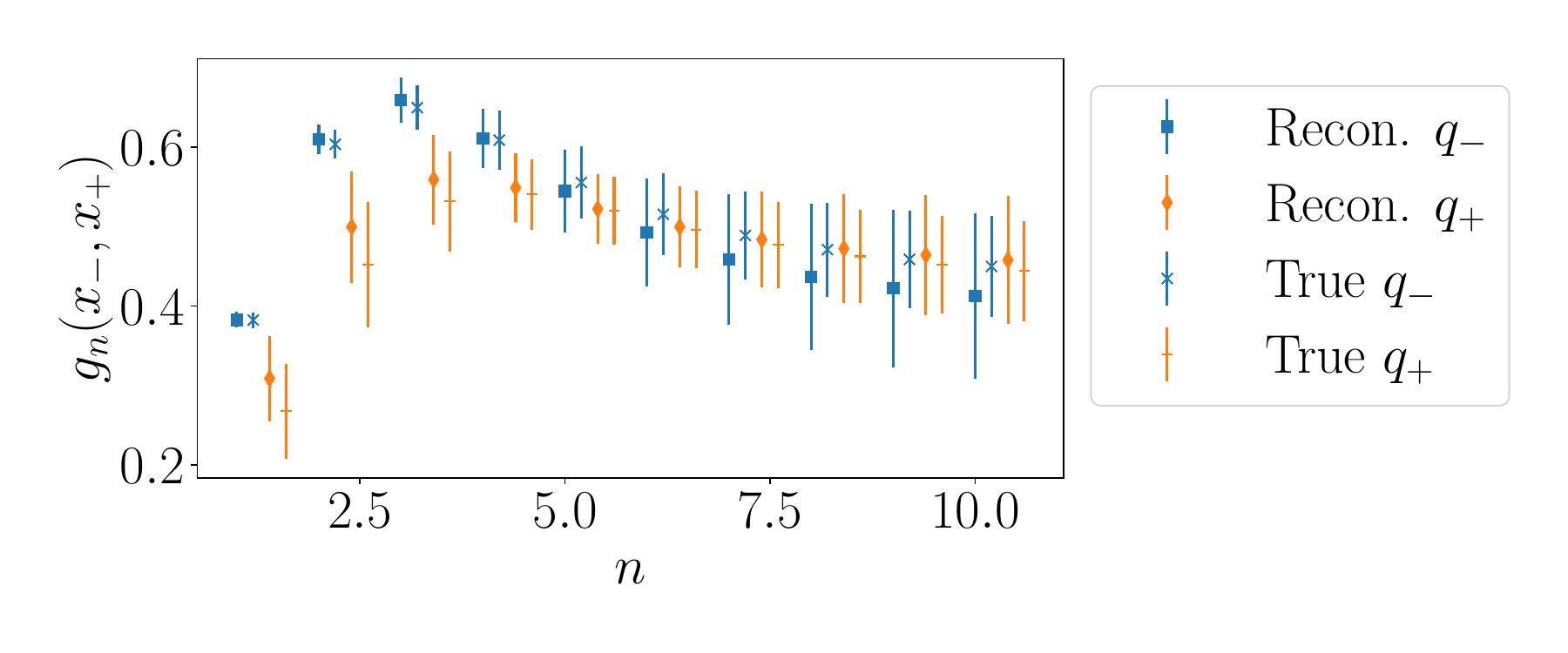}
    \includegraphics[width=\linewidth]{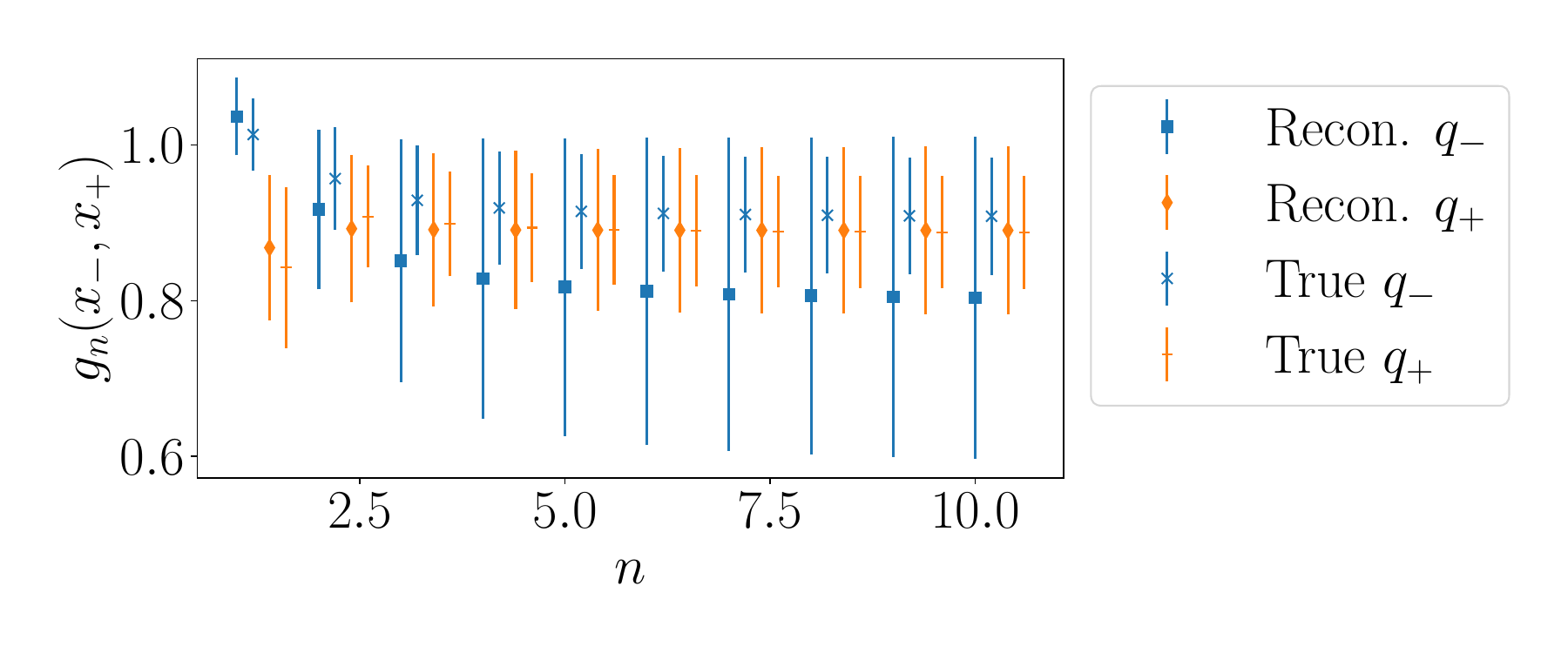}
    \includegraphics[width=\linewidth]{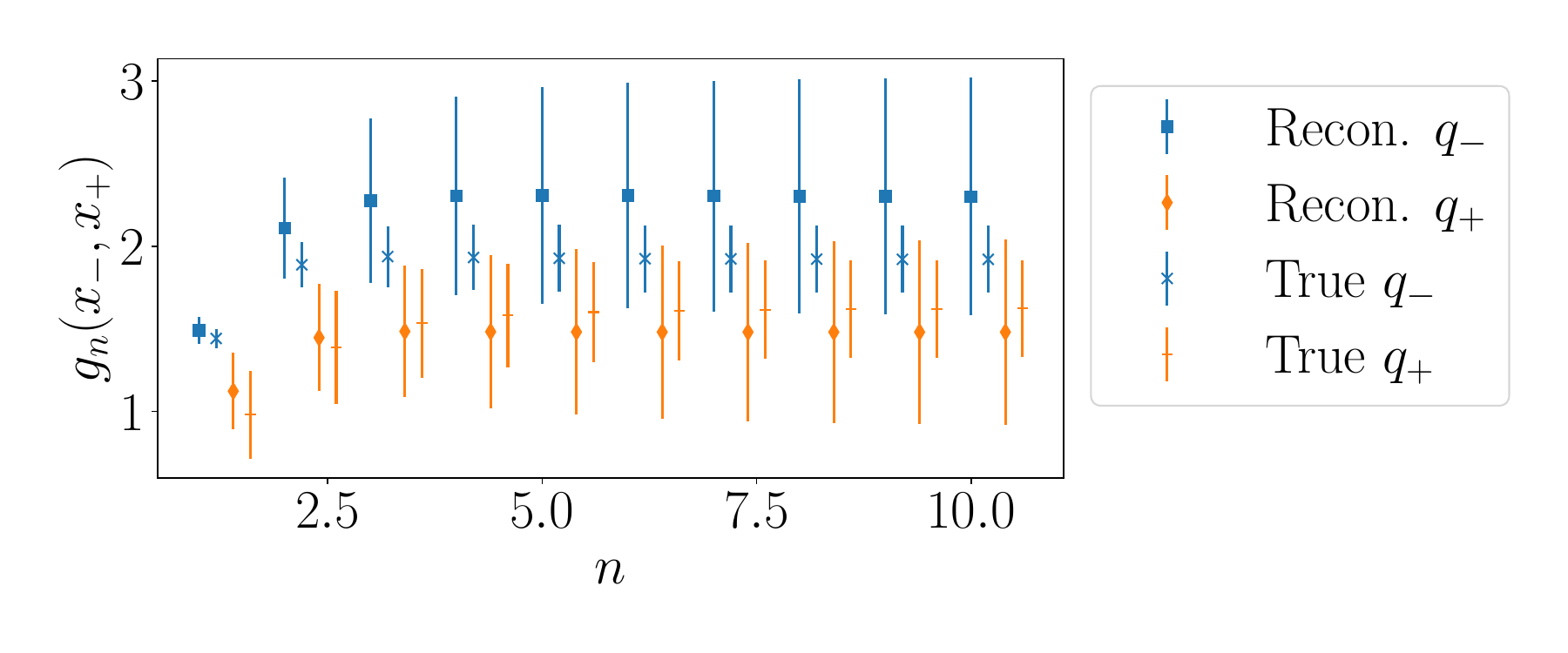}
    \caption{The Gaussian windows with, in descending order, $(x_-,x_+)$ = (0.1, 1), (0.3, 0.5), and (0.1, 0.3). }\label{fig:other_gauss_windows}
\end{figure}

This section shows the window observables for other choices of the window, $(x_-,x_+)$ = (0.1, 1), (0.3, 0.5), and (0.1, 0.3) . Fig.~\ref{fig:other_window_moms} shows the window moments and Fig~\ref{fig:other_gauss_windows} shows the Gaussian windows. Similar to the window chosen for the main text, these show consistent reproduction from both real and imaginary component. At larger $n$, the error estimated from the synthetic data is overestimated while smaller $n$ tends to have more precise reproduction. 

\bibliography{window.bib}

\end{document}